\documentclass{tcibook}
\usepackage{fancyhea}
\usepackage{work}
\usepackage{bm}       
\usepackage{graphicx}
\usepackage{hyperref}      
\usepackage{lineno}

\newcommand{\nc}{\newcommand}  

\def\beq{\begin{equation}}
\def\eeq#1{\label{#1}\end{equation}}
\def\eeqn{\end{equation}}

\newenvironment{Eqnarray}%
   {\arraycolsep 0.14em\begin{eqnarray}}{\end{eqnarray}}
\def\beqa{\begin{Eqnarray}}
\def\eeqa#1{\label{#1}\end{Eqnarray}}
\def\eeqan{\end{Eqnarray}}

\nc{\ra}{\rightarrow}  
\nc{\slsh}{\slash\hspace*{-0.22cm}}
\def\Re{{\cal R \mskip-4mu \lower.1ex \hbox{\it e}\,}}
\def\Im{{\cal I \mskip-5mu \lower.1ex \hbox{\it m}\,}}

\nc{\vev}[1]{ \left\langle {#1} \right\rangle }
\nc{\bra}[1]{ \langle {#1} | }
\nc{\ket}[1]{ | {#1} \rangle }
\nc{\fb}{\,{\rm fb}^{-1}}
\nc{\ev}{{\rm eV}}
\nc{\kev}{{\rm keV}}
\nc{\Mev}{{\rm MeV}}
\nc{\gev}{{\rm GeV}}
\nc{\tev}{{\rm TeV}}
\nc{\mev}{{\rm MeV}}

\def\del{\partial}
\def\Dslash{\not{\hbox{\kern-4pt $D$}}}
\def\dslash{\not{\hbox{\kern-2pt $\del$}}}
\def\pslash{\not{\hbox{\kern-2pt $p$}}}
\def\ETmiss{ \not{\hbox{\kern-4pt $E$}}_T }

\def\ee{e^+e^-}

\def\sstw{\sin^2\theta_W}

\def\msb{{\bar{\ssstyle M \kern -1pt S}}}

\begin{document}

\def\bibname{References}
\bibliographystyle{plain}

\raggedbottom

\pagenumbering{roman}

\parindent=0pt
\parskip=8pt
\setlength{\evensidemargin}{0pt}
\setlength{\oddsidemargin}{0pt}
\setlength{\marginparsep}{0.0in}
\setlength{\marginparwidth}{0.0in}
\marginparpush=0pt


\renewcommand{\chapname}{chap:intro_}
\renewcommand{\chapterdir}{.}
\renewcommand{\arraystretch}{1.25}
\addtolength{\arraycolsep}{-3pt}

\thispagestyle{empty}
\begin{centering}
\vfill

{\Huge\bf Planning the Future of U.S. Particle Physics}

{\Large \bf Report of the 2013 Community Summer Study}

\vfill

{\Huge \bf Chapter 1:  Summary}

\vspace*{2.0cm}
{\Large \bf Corresponding Author:  J. Rosner}

\vfill

{\large  Study Conveners: M. Bardeen, W. Barletta, L.~A.~T.~Bauerdick, R. Brock,
D.~Cronin-Hennessy, M.~Demarteau, M.~Dine, J.~L. Feng, M. Gilchriese,
S. Gottlieb, J.~L.~Hewett, R. Lipton, H.~Nicholson, M.~E. Peskin,
S. Ritz, I.~Shipsey, H. Weerts}\\
\vspace{1cm}

{\large Division of Particles and Fields Officers in 2013:
J.~L. Rosner (chair), 
I. Shipsey (chair-elect), 
N. Hadley (vice-chair),
P. Ramond (past chair)}\\
\vspace{1cm}

{\large Editorial Committee:
R.~H. Bernstein,
N. Graf,
P. McBride,
M.~E. Peskin,
J.~L. Rosner,
N.~Varelas,
K. Yurkewicz}

\vfill

\end{centering}

\newpage


\mbox{\null}

\vspace{3.0cm}

{\Large \bf Authors of Chapter 1:}

\vspace{2.0cm}

{\bf J. L. Rosner}, M. Bardeen, W. Barletta, L. A. T.  Bauerdick, R. H. Bernstein, R. Brock,
D. Cronin-Hennessy, M.~Demarteau, M.~Dine, J.~L.~Feng, M. Gilchriese,
S. Gottlieb, N. Graf, N. Hadley,  J. L. Hewett, R. Lipton, P. McBride,
H.~Nicholson, M.~E.~Peskin, P. Ramond, S.~Ritz, I. Shipsey, N. Varelas, H. Weerts,
K. Yurkewicz
 
\newpage


\begin{center}
{\Large\bf Foreword}
\end{center}

Particle physics research in the United States is a vibrant field, with
discoveries stretching back more than half a century. Over the past year,
more than 1000 U.S. particle physicists engaged in an in-depth process to
define the most important questions for our field and identify the most
promising opportunities to address these questions. 

The process had its roots in a series of meetings held at Snowmass, Colorado
over the past thirty years to take stock of progress in particle physics and
chart the field's future. The last such periodic meeting was held in 2001.
In 2011 the Division of Particles and Fields of the American Physical
Society decided that the time was ripe for a new community study.
Preparatory meetings during 2012 and 2013 began with a Community Planning
Meeting at Fermilab, October 11-13, 2012. A final plenary meeting was held
at the University of Minnesota, July 29 - August 6, 2013.

The 2013 Community Summer Study (the ``Snowmass study'') was
designed to enumerate the questions the particle physics community wishes
to answer over the next two decades, and plan how to answer them.  The study
does not prioritize activities, but aims to ask and answer hard questions.
A subsequent prioritization panel with broad community representation will
place these questions and answers within realistic budgetary scenarios.  The
study has produced this resource book, charting aspirations of the U.S.
particle physics community over the next ten to twenty years, for use by the
panel in its deliberations.  We also intend the report to convey the
health and diversity of the U.S. program, in a global context, to our
colleagues and fellow citizens.

Although we found it convenient to retain the ``frontier'' categories of the
previous Particle Physics Project Prioritization Panel (``P5''), whose last
report was issued in 2008, the division of the field into such categories
should not obscure the focus on fundamental questions of physics, which, by
their nature, cross such frontiers.  These inter-frontier discussions have
been a major component of the Minnesota meeting.

This volume can be read at various levels.  An executive summary precedes a
more detailed summary chapter.  Each frontier (Intensity, Energy, Cosmic,
Theory, Capabilities, Instrumentation, Computing, and Communication) has
its own chapter containing further details.  Reference is made to
submissions by each frontier's subgroups, and to contributed white papers.

We thank all the host institutions and organizers of the preparatory and
Minnesota meetings for their efforts on behalf of this review of our field.

For the Conveners and the DPF Executive Committee,
\begin{center}
\begin{tabular}{l l}
Jonathan L. Rosner & DPF Chair \\
Ian Shipsey        & DPF Chair-Elect \\
Nick Hadley        & DPF Vice-Chair \\
Pierre Ramond      & DPF Past Chair \\
\end{tabular}
\end{center}

\newpage
\begin{center}
{\Large\bf Executive Summary}
\end{center}

The 2013 Community Summer Study, known as ``Snowmass,'' brought together nearly
700 physicists to identify the critical research directions for the United
States particle physics program. Commissioned by the American Physical Society,
this meeting was the culmination of intense work over the past year by more
than 1000 physicists that defined the most important questions for this field
and identified the most promising opportunities to address them. This Snowmass
study report is a key resource for setting priorities in particle physics.

Particle physicists seek to understand the structure of the universe.  We
address two main questions:  What are the most elementary constituents of
nature, and what are the forces that cause them to interact? These questions are
fundamental, and the desire to explore them is a defining characteristic
of the human spirit. At the same time, finding the answers has practical value:
It helps drive technical innovation in instrumentation, computing, and
accelerators, and leads to the development of a skilled technical workforce.  The
development of new technologies, from industrial techniques, to medical
imaging, high-performance computing, and beyond, has continually improved the
quality of human life.

The discovery of the Higgs boson in 2012 was a remarkable achievement made
possible by decades of worldwide collaboration.  The existence of the
Higgs
boson was predicted in the
1960s.  By last year it was the sole missing piece of the theory we call the
Standard Model.  This theory provides a coherent picture of the strong, weak,
and electromagnetic interactions, with the latter two unified in an
``electroweak'' theory.  The Standard Model contains quarks, leptons, force
carriers, and now the Higgs boson.

However, the Standard Model still leaves significant questions unanswered. What
is the nature of the Higgs boson?  What can we learn from discovering that
neutrinos have mass?  Can the known forces be further unified?  The particles
of the Standard Model make up only 5\% of the universe --- what is the other
95\%?  Why is the universe almost all matter and no antimatter?

Many different techniques are needed to answer these questions.  Particle
physics explores three basic frontiers, often called cosmic, energy, and
intensity.  We continue that designation in this report. Each frontier uses
its own tools and techniques, but all ultimately address the same fundamental
questions.

The experiments that address these questions are ambitious, large-scale
projects.  Mounting them requires long-term vision.  We are fortunate that
our priorities are shared by physicists in other regions of the world, so that
these experiments can be realized as global partnerships.  The U.S. brings
crucial leadership, design talent, technology, and resources that will be
essential to these experiments wherever they are located.

The outline that follows introduces the future directions necessary for further
progress in our understanding of nature at its most fundamental level.  The
order does not reflect prioritization.
\medskip

{\noindent\Large\bf\it Particle Physics Frontiers}

\noindent{\bf Intensity Frontier:} Experiments at the ``Intensity Frontier''
explore fundamental questions by using precision measurements
to search for extremely rare processes and for tiny deviations from
expectations.  They can reveal new laws of physics at very high energies, in
many cases exploring beyond the direct reach of high-energy accelerators.  They
often require the greatest possible beam intensities and ultra-sensitive
detectors.  The Snowmass study identified facilities and experiments that will
ensure the global leadership of the U.S. in Intensity Frontier science.

Neutrinos are the most elusive of the known fundamental particles.  We know of
three species, which can change (``oscillate'') into one another under the
right conditions.  We have recently discovered that one of the neutrino
oscillation parameters is large enough to let us measure two fundamental
properties of neutrinos.  First, we can hope to determine which of the three
neutrinos is heaviest and which is lightest (the mass hierarchy).  Second, we
can determine if leptons violate CP (charge-parity) symmetry, one of the most
important symmetries of nature.   We have known for almost fifty years that CP is
violated by quarks; now the focus has shifted to leptons.

The Long-Baseline Neutrino Experiment (LBNE) will measure the mass hierarchy
and is uniquely positioned to determine whether leptons violate CP.
Future multi-megawatt beams aimed at LBNE, such as those 
from Project X at Fermilab, would enable studies of CP violation in neutrino
oscillations with conclusive accuracy.  An underground LBNE detector would
also permit the study of atmospheric neutrinos, proton decay, and precision
measurement of any galactic supernova explosion.  This represents a vibrant
global program with the U.S. as host.

The Snowmass study discussed further opportunities to learn about neutrinos. An
upgrade of the IceCube experiment, involving photodetectors buried in Antarctic
ice, could provide a promising approach to measure the mass hierarchy using
atmospheric neutrinos.  Next-generation neutrinoless double-beta decay
experiments could reach the sensitivity necessary to determine whether
neutrinos are their own antiparticles, and are a critical component of a strong
neutrino program.

Transitions among different types, or ``flavors,'' of quarks and leptons
provide essential probes of new physics, and are a central element of the
Intensity Frontier.  Substantial progress toward understanding these
transitions can be made in this decade with experiments utilizing the Fermilab
accelerator complex.  These include a new measurement of the anomalous magnetic
moment of the muon, a sensitive search for muon--to--electron conversion, and a
proposed experiment to probe rare $K$ meson decays to a new level of precision.

If particles have a property known as electric dipole moments (EDMs), they
will violate CP symmetry.  Finding EDMs would have
profound implications, allowing us to check models of CP violation and to
help understand the preponderance of matter in the universe.  Project X could
provide the means to make incisive EDM measurements of unprecedented precision.

The U.S. has also made substantial investments in other flavor physics
experiments.  Snowmass studies showed that the U.S. can capitalize on them,
achieving unprecedented sensitivities with the capabilities offered by the LHCb
experiment at CERN, Belle-II and J-PARC in Japan, and BESIII in China.  Each
probes a different aspect of new physics in a unique way.  Project X will
complement and extend results of these experiments.

New light, weakly coupled particles appear in many theoretical models,
especially those of the cosmic dark matter.  Some searches for these particles
are feasible with intense beams and comparatively modest detectors at existing
facilities.  The Snowmass studies identified a rich, diverse, and low-cost
program with a potential for high-impact discoveries, illustrating the
importance of modest-scale experiments to complement large-scale efforts.

\noindent{\bf Energy Frontier:} Experiments at the ``Energy Frontier'' make
use of high-energy colliders to directly produce heavy elementary particles and
to search for new ones. The properties of the Higgs boson raise crucial
questions that guide large parts of the future particle physics program. The
Higgs boson discovery calls for a three-pronged research program at high-energy
accelerators, to: (1) determine the properties of the Higgs boson as accurately
as possible, (2) make precise measurements of the heavy particles $W$, $Z$,
and the top quark, which can carry the imprint of the Higgs field, and (3)
search for new particles predicted by models of the Higgs boson
and electroweak symmetry breaking.  These topics also overlap with those in
other frontiers.  The expectation of particles with masses at or above one
TeV directly motivates searches for dark matter particles and flavor-changing
rare decays.

For at least the next fifteen years, the experiments at the Large Hadron
Collider (LHC) at CERN will drive the Energy Frontier program forward.  The LHC
experiments are successful global collaborations.  U.S. leadership and
contributions to detector and accelerator components, technology, and physics
insight have been indispensable to that success.

Following the Higgs boson discovery, the LHC moves into a phase of precision
study of the properties of this particle.  The high-luminosity LHC will measure
Higgs boson interactions at the few-percent level.  LHC operation at higher
energy over the next ten years, and, later, at higher luminosity, will continue
the search for new particles produced through the strong or electroweak
interactions.  It will probe for new dynamics of the $W$, $Z$, and Higgs bosons
at TeV energies and study rare decays using billions of top quarks.

Compelling science motivates continuing this program with experiments at lepton
colliders.  Experiments at such colliders can reach sub-percent precision in
Higgs boson properties in a unique, model-independent way, enabling discovery
of percent-level deviations from the Standard Model predicted in many theories.
They can improve the precision of our knowledge of the $W$, $Z$, and top quark
well enough to allow the discovery of
predicted new-physics effects.  They search for new particles in a manner
complementing new particle searches at the LHC.  A global effort has
completed the technical design of the International Linear Collider (ILC)
accelerator and detectors that will provide these capabilities in the latter
part of the next decade.  The Japanese particle physics community has declared
this facility as its first priority for new initiatives.

The Snowmass study considered many other options for high-energy colliders that
might be realized over a longer term.  These included higher-energy linear
colliders, circular $e^+e^-$ colliders, muon colliders, and photon colliders,
all of which merit continued study.  The Snowmass study identified, in
particular, the promise of a 100 TeV-class hadron collider (VLHC), which would
provide a large step in energy with great potential for new insights into
electroweak symmetry breaking and dark matter. The feasibility of such a
machine should be clarified through renewed accelerator R\&D and physics
studies over the next decade.

In all of the above projects, U.S. leadership in developing detector and
accelerator technologies is playing a critical role. These U.S. initiatives
are essential to meet the world-wide scientific goals in particle physics.

\noindent{\bf Cosmic Frontier:} 
We now know that 95\% of the universe is in the form of dark matter and dark
energy.  These components are responsible for the structures and accelerated
expansion of the universe observed today, but their fundamental nature is
almost completely unknown.  ``Cosmic Frontier'' experiments are designed
to determine the nature of dark matter and dark energy and to use the universe
as a laboratory to search for new fundamental particles and interactions.
Along with the other frontiers, the Cosmic Frontier provides particle physics
with clear evidence for physics beyond the Standard Model, profound questions
of popular interest, frequent new results and surprises with broad impacts,
many opportunities for discovery with unique probes, important cross-frontier
topics, and a full range of project scales providing flexible programmatic
options.

Our studies showed how different approaches to dark matter --- direct
detection; indirect detection of gamma rays, neutrinos, and antimatter;
accelerator-based searches; simulations; and
astrophysical surveys --- provide unique and necessary information.  A census
of current and proposed experiments and their capabilities shows that new
direct and indirect detection experiments will probe dark matter
masses inaccessible to colliders and provide leaps in sensitivity at
moderate cost.  For many leading dark matter candidates,
including well-known examples of weakly-interacting massive particles
and axions, this progress will lead to the first non-gravitational
signals of the dark universe and open the door to the identification
of dark matter, with far-reaching implications for both particle
physics and astronomy.

Snowmass studies also strongly reinforced the roles that cosmic surveys play
in particle physics.  Current and upcoming dark energy imaging and
spectroscopic surveys will shrink the errors, as recommended in
previous community studies, but they will do even more for particle
physics. The richness of the data and detailed attention to the
reduction of systematic uncertainties will enable many new tests of
the behavior of dark energy and general relativity over a wide range
of
cosmic 
distance and time scales.

The universe appears to have undergone an enormous expansion in the very first
moments after the Big Bang.  This phenomenon, known as inflation, helps us
understand many subsequent details of how the universe developed.  Cosmic
microwave background (CMB) experiments will probe the physics of inflation
with enough sensitivity to test many of our ideas about the birth of the
universe.

Cosmic Frontier experiments can also help us understand neutrinos.  Studies of
the CMB, measurements of the distribution and apparent shapes of galaxies, and
polar-ice neutrino experiments will provide precise information about
neutrinos, including the mass hierarchy, the sum of their masses, and the
number of light neutrinos.  These experiments provide information complementary
to laboratory studies of neutrinos, and both types combine to create a powerful
means for detailed neutrino investigations.

Finally, the Snowmass process reiterated the unique information we can
gain from studies of cosmic particles and the detection of significant
numbers of the highest-energy cosmic rays produced in nature. These studies
provide a window on proton interactions at energies beyond the reach
of the
LHC and include
the detection of extremely high-energy neutrinos produced in the interactions
of cosmic rays with CMB photons, which will enable the study of neutrino
interactions at center-of-mass energies up to 100 TeV.

For all of these areas, the Snowmass study identified essential
technologies and facilities, the advances required in theoretical
models, and experiments with great promise.  The largest projects are,
appropriately and necessarily, international.  The U.S. is still the
leader in this quickly evolving area, but other regions with intensive
interest in this physics are advancing rapidly.

\noindent{\bf Theoretical Physics:} 
Progress in science is based on the interplay between theory and experiment,
between having an idea about nature and testing that idea in the laboratory.
Neither can move forward without the other.  The U.S. has been a world leader
for many decades in particle theory, and a sustained strong and vibrant program
remains essential for the success of U.S. particle physics. Theoretical
physicists have been a driving force in both the development and testing of
the Standard Model, including the discovery of the Higgs boson.  They play
crucial roles in formulating the big questions in the field, setting out
hypotheses that address them, and proposing experimental strategies to confirm
or refute them.  At the same time, theorists seek new structures that might
provide unanticipated results.  In contrast to experiment, theory depends less
on facilities and equipment; the success of the U.S. program rests mainly on
principal investigators in universities and national labs, working with
postdoctoral fellows and graduate students, and collaborating with both U.S.
and international researchers.
\medskip

{\noindent\Large\bf\it Enabling Frontiers}

\noindent{\bf Accelerator Capabilities:}
Accelerator-based experiments continue to be the mainstay of transformational
physics on both the Energy and Intensity Frontiers.  Progress in these
frontiers demands advancing the capabilities of accelerator facilities.
The U.S. has been a leader in accelerator physics, its critical and supporting
technologies, and the innovative design of research facilities.  Its continuing
leadership is necessary for future discoveries.

The LHC incorporates major U.S. contributions, including high-field
superconducting magnets that focus the beams into collision.  For future
Energy Frontier facilities, U.S. laboratories have pioneered the technology
of Nb$_3$Sn magnets that will permit higher fields than possible with the
present LHC technology. The U.S. LHC Accelerator Research Program (LARP) gives
the U.S. a world-leading ability to develop high-field superconducting
accelerator magnets --- a capability central to the LHC luminosity upgrade
and for a future proton collider with far greater energy than the LHC.  A 100
TeV-class hadron collider (VLHC) is within the development reach of existing
materials for a tunnel of 100 km or larger.  A collider at even higher energy
may require new classes of superconducting magnets and novel ways of handling
synchrotron radiation.

The ILC, as described in its Technical Design Report, is ready to proceed to
construction.  Its design incorporates U.S. contributions in accelerator
theory, damping ring design, superconducting accelerator
technology, and beam control and delivery.  Concepts for multi-TeV
lepton colliders include the CLIC two-beam accelerator, plasma wakefield
accelerators driven either by beams or lasers, and a muon collider.

Accelerators for proposed Intensity Frontier experiments need to deliver
multi-megawatt proton beams with flexible,
experiment-dependent timing structures --- demands beyond the capabilities of
any existing accelerator. Multi-megawatt beams are the focus of vigorous R\&D
for both the Project X superconducting linac and DAE$\delta$ALUS cyclotrons.
A broad range of highly flexible timing structures is being pursued within
Project X.

Managing huge stored energy and controlling beam losses to minuscule levels
will be essential to operation of all frontier physics accelerators.
Specific challenges include generating high-quality beams, modeling beam
dynamics, and managing material damage in high-power targets.
Maximizing the potential and minimizing the risks in future accelerator
projects requires integrated multi-laboratory programs to increase the
readiness of critical technologies.  Yet engineering-intensive programs must
not squeeze out visionary, innovative research in basic accelerator science.

A strong U.S. research program of accelerator stewardship benefits all
areas of science and industry that use accelerator technology.  The broad
application of superconducting technology to accelerator-based science
exemplifies the payoff of long-range investment that transcends individual
projects.  Investment in national laboratories and research universities,
with their broad expertise and technical infrastructure, will yield new
generations of accelerators capable of higher energies, more intense beams,
and more efficient operation.

\noindent{\bf Underground Laboratory Capabilities:}
Many experiments searching for dark matter, proton decay, or seeking to
determine the properties of neutrinos must be located underground to shield the
sensitive experiments from cosmic ray backgrounds. Underground facilities are
located or proposed in North and South America, Europe, Asia and in the
Antarctic ice. The scope of underground capabilities in all regions is expected
to increase by the end of the decade to accommodate the experimental demand.
Locating LBNE underground would allow this experiment to realize its full
scientific potential and could make it an anchor for a future domestic
underground laboratory, bringing to the U.S. a broad range of compelling
experiments and researchers from around the world.

\noindent{\bf Instrumentation:} Instrumentation enables experiments
to answer the science questions facing particle physics. Particle physics has
a long and distinguished history of inventing, designing, and building the
specialized instrumentation required for its experimental research.  The
physics requirements of many experiments in the Energy, Intensity and Cosmic
Frontiers entail very large-scale detectors, but the cost involved in simply
scaling up existing technologies is becoming prohibitive. In order for the
field to stay competitive, new innovative technologies will need to be
developed.  Investment in the development of these new technologies is not a
luxury but a prerequisite for the sustainability of the field. It has therefore
become very important for the particle physics community to establish a
mechanism for developing and implementing a coherent vision for the future
direction for particle physics instrumentation.  

The Snowmass study has formulated a vision for a U.S. instrumentation program
for particle physics.  We identified key barriers to answering the science
questions and recognized select technologies for further investment 
enabling the U.S. to retain a leadership position in a broad global
experimental program.

Accelerator experiments will require fast, radiation-hard, low-mass, highly
segmented, intelligent and sophisticated trackers and vertex detectors;
cost-effective, highly segmented, radiation-hard calorimeters with excellent
energy resolution; and high-speed data acquisition systems.  Experiments
studying particle interactions with small cross sections or rare decays
typically need cost-effective large volume detectors with efficient background
rejection using large area charged particle detection and/or optical readout
systems.  Some of these experiments need materials and sensors with ultra-low
intrinsic radioactivity while others need high-granularity detectors with fast
timing and good energy resolution.  Many experiments in the cosmic frontier
utilize large arrays of ultra-low-noise electromagnetic radiation detectors.

Executing a strategy to develop innovative technologies requires integrating
the diverse capabilities and resources of universities, national laboratories,
other branches of science, and industry into detector R\&D collaborations, 
emphasizing the importance of innovation through a domestic instrumentation
development program.  A coordinating panel for advanced detectors would help
articulate the mission of this program and facilitate its implementation.  The
goals of the program are to develop both incremental and transformational
cost-effective technologies with maximal scientific reach, based on the
technological strengths in the U.S. A stable and adequately funded generic
instrumentation program will ensure that particle physics invests in its future
and establishes a foundation for a competitive, healthy long-term program.

\noindent{\bf Computing:}
Computing is essential to all particle physics experiments and
in many areas of theoretical physics.  Although some hardware is customized,
most of it comes from commercial vendors.  Thus, selecting the right hardware
and using it efficiently are essential to achieve our scientific goals.

We have to train the personnel required to develop and support the parallel
programs needed now and in the future.  Increased parallelism is required
because
of changes in chip technology and the evolution of high-performance systems to
include multi-core chips and accelerators.  It is also important to support the
development and implementation of new algorithms in several theoretical areas.

Particle physicists should help in the planning of U.S. and
international networks, as LHC
upgrades will place more demands on the distributed computing systems for ATLAS
and CMS.  Failure to do so could lead to bottlenecks from wide-area networks,
hampering the analysis of results from those and future experiments.  Funding
agencies should encourage enhanced coordination of software
development across all frontiers.  The needs of Intensity Frontier experiments
are not yet at the level of LHC experiments, but will grow.

Future experiments will demand better storage capacity and bandwidth.  Disk
prices may not drop as rapidly as they have in the past.  Scientists working
on the cutting edge of computing must also continue to detail their needs to
the national supercomputing centers.  The funding agencies should seek
community input on the appropriate mix of resources dedicated to a particular
project and shared computing resources.

Early attention to these issues can increase efficiency,
reduce costs, enable significantly more realistic theoretical calculations,
and avoid computing bottlenecks that could limit scientific progress.
\medskip

\noindent{\Large\bf\it Communication, Education, and Outreach}

The particle physics community recognizes the critical importance of consistent
and coherent communication, education, and public outreach (CE\&O).  These
foster nationwide support for the field and develop the next generation of
scientists, engineers, and scientifically literate citizens.  More of us should
engage in CE\&O activities to translate the American public's fascination with
particle physics research into the support necessary to enable the field to
answer its biggest questions.  Existing activities should be augmented with
dedicated personnel who will enhance these efforts, provide nationwide
coordination, and spearhead new initiatives.
These personnel would develop materials for use in informing the public
about direct and indirect applications of particle physics research. They
would put in place sustainable methods to collect statistics on workforce
development and technology transfer. They could provide professional
development opportunities for educators, as well as creating new learning
opportunities for students of all ages.
\newpage

\noindent{\Large\bf\it Conclusion}

With the completion of the Standard Model, particle physicists now turn their
attention to still deeper questions about the nature of matter and the
constituents of the universe.
This report proposes an ambitious array of new experiments.  We
consider it realistic to carry out these experiments through a long-term plan
and through global partnerships.  Particle physicists have been the pioneers of
large-scale scientific projects.  We have constructed facilities of
unprecedented scale, including the Tevatron and the Large Hadron Collider,
through decades-long programs requiring world-wide collaboration.  These led
to discoveries that are the foundation of our current success.

Several strategic goals have emerged from the Snowmass study.

\begin{itemize}

\item Probe the highest possible energies and distance scales with the
existing and upgraded Large Hadron Collider and reach for even higher
precision with a lepton collider; study the properties of the Higgs boson
in full detail.

\item Develop technologies for the long-term future to build multi-TeV
lepton colliders and 100 TeV hadron colliders.

\item Execute a program with the U.S. as host that provides precision tests
of the neutrino sector with an underground detector; search for new physics
in quark and lepton decays in conjunction with precision measurements of
electric dipole and anomalous magnetic moments.

\item Identify the particles that make up dark matter through complementary
experiments deep underground, on the Earth's surface, and in space, and
determine the properties of the dark sector.

\item Map the evolution of the universe to reveal the origin of cosmic
inflation, unravel the mystery of dark energy, and determine the ultimate
fate of the cosmos.

\item Invest in the development of new, enabling instrumentation and
accelerator technology.

\item Invest in advanced computing technology and programming expertise
essential to both experiment and theory.

\item Carry on theoretical work in support of experimental projects and to
explore new unifying frameworks.

\item Invest in the training of physicists to develop the most creative
minds to generate new ideas in theory and experiment that advance science
and benefit the broader society.

\item Establish a nationally coordinated communication, education and outreach 
effort, supported by a dedicated team, to convey the excitement and value of our field to others.

\end{itemize}

In pursuit of these projects, we have developed a community that links
together scientists from all regions of the world pursuing common goals.  Our
community is ready and eager to carry out the next steps in humankind's
quest to understand the basic workings of the universe.

\newpage

\tableofcontents

\newpage

\mbox{\null}

\newpage

\pagenumbering{arabic}

\chapter{Summary of the 2013 Community Summer Study}

\bigskip

\section{Introduction}
\label{sec:intro}

The 2013 Community Summer Study, known as ``Snowmass,'' sought to identify
the critical research directions for the United States particle physics
program.  This meeting was the culmination of intense work over the past year to define
the most important questions for this field and identify the most
promising opportunities to address them.  The resulting study report,
presented here, is a key resource for setting priorities in particle
physics.

Through the previous six decades of precision and discovery-level particle
physics, we have learned much about the basic laws that govern the Universe.
We have uncovered the laws that describe the subnuclear forces and, with the
discovery of the Higgs boson, the agent that we believe should give mass to all
elementary particles.  However, there is still much that we do not understand.
The advances in our knowledge of elementary particle physics have sharpened the
questions in that domain.  Recent discoveries about the matter and energy
content of the Universe have brought new questions that are equally
fundamental, and equally vexing.

One of the goals of Snowmass was to develop a framework of scientific questions
that can form the basis for a future program in particle physics, and
to survey experiments that would address these questions.  To introduce
a summary of the results of Snowmass, we propose a basic set of
questions about particle physics 
whose answers are not yet known.   The search for these
 answers will be carried out with a broad
range of experimental methods, cutting across the frontiers around which the
Snowmass study was organized.

\begin{enumerate}

\item{How do we understand the Higgs boson?  What principle determines its
couplings to quarks and leptons?  Why does it condense and acquire a vacuum
value throughout the Universe?  Is there one Higgs particle or many?  Is the
Higgs particle elementary or composite?}

\item{What principle determines the masses and mixings of quarks and leptons?
Why is the mixing pattern apparently different for quarks and leptons?
Why is there CP violation in quark mixing?  Do leptons violate CP?}

\item{Why are neutrinos so light compared to other matter particles? Are
neutrinos their own antiparticles? Are their small masses connected to the
presence of a very high mass scale?  Are there new interactions that are invisible
except through their role in neutrino physics?}

\item{What mechanism produced the excess of matter over anti-matter that we see
in the Universe? Why are the interactions of particles and antiparticles not
exactly mirror opposites?}

\item{Dark matter is the dominant component of mass in the Universe.  What is
the dark matter made of?  Is it composed of one type of new particle or
several?  What principle determined the current density of dark matter in the
Universe?  Are the dark matter particles connected to the particles of the
Standard Model, or are they part of an entirely new dark sector of particles?}

\item{What is dark energy?  Is it a static energy per unit volume of the
vacuum, or is it dynamical and evolving with the Universe?  What principle
determines its value?}

\item{What did the Universe look like in its earliest moments, and how did it
evolve to contain the structures we observe today? The inflationary Universe
model requires new fields active in the early Universe. Where did these come
from, and how can we probe them today?}

\item{Are there additional forces that we have not yet observed? Are there
additional quantum numbers associated with new fundamental symmetries? Are the
four known forces unified at very short distances?  What principles are
involved in this unification?}

\item  Are there new particles at the TeV energy scale?  Such particles are
motivated by the problem of the Higgs boson, and by ideas about space-time
symmetry such as supersymmetry and extra dimensions. If they exist, how do
they acquire mass, and what is their mass spectrum? Do they provide new sources
of quark and lepton mixing and CP violation?

\item  Are there new particles that are light and extremely weakly interacting?
Such particles are motivated by many issues, including the strong CP problem,
dark matter, dark energy, inflation, and attempts to unify the microscopic
forces with gravity.  What experiments can be used to find evidence for these
particles?

\item Are there extremely massive particles to which we can only couple
indirectly at currently accessible energies? Examples of such particles are
seesaw heavy neutrinos or grand unified scale particles mediating proton decay.  How
can we demonstrate that these particles exist?

\end{enumerate}

The search for answers to these questions is intimately tied to the development
of technology.  Particle physics experiments and accelerators put
extraordinary demands on sensors, precision engineering, and data management,
incorporated into devices of very large scale.  Our community invents new
technologies to address these needs and develops them into usable form.  The
progress of our field requires both technology development directed at the 
problems of specific experiments and the development of new technologies that
provide higher performance or decreased cost for devices with broad
application.  This technology development for accelerators and detectors
ultimately benefits all of physical science.

In many areas of physics experimentation, there are specific technological
developments that would be of enormous benefit. Existing technologies are
unlikely to meet the science needs of future particle physics experiments.
New technologies need to be explored that could lead to transformative
advances, enabling cost-effective particle physics experiments but also new
initiatives of broad importance.

In developing such technoloigies, we need to address several questions.

\begin{enumerate}

\item{Experiments continue to reach for rarer processes, more precise
measurements, higher energies and luminosities, and more inclusive observations.
How do we achieve the finer granularity, larger volume, more radiation-hard,
lower-cost, and higher-speed detectors that will in large part determine our
experimental reach?}

\item{Paradigm-altering technology developments are occurring in electronics
and materials design, potentially offering breakthrough capabilities.  How can
these advances be incorporated into new detectors with improved overall
performance? How do we make best use of the resources available in
universities, national laboratories, and industry to develop new detector
systems?}

\item{What technologies will be needed to acquire, analyze, and store the
enormous amounts of data from future experiments?  Can local intelligence be
incorporated to manage data flow?  How will we fully and efficiently utilize
data stored in large databases?}

\item{Scaling of current accelerator designs to higher energy leads to machines
of very large size, cost, and power demand.  Can new technologies lead to more
practical strategies?  Is there an ultimate highest energy for colliders?}

\item{Proposed experiments at a range of energy scales call for particle
beams of extreme brightness.  Are there technologies to achieve high beam
power in a better-controlled and more cost-effective way?}

\end{enumerate}

It is important for particle physicists to share the excitement and benefits
of our field with a broader public.  To that end:

\begin{enumerate}

\item{How do we engage particle physicists in communication, education and
outreach activities so as to convince policy makers and the public that
particle physics is exciting and worth supporting?}

\item{How do we educate a talented and diverse group of students who
    choose to enter
particle physics and other STEM careers, including science teaching?}

\end{enumerate}

In the following chapters, we discuss these issues in more detail and
explain
how they will be addressed in the coming decades by  new 
initiatives in  particle physics.  The discussion is organized along
the lines of the physics frontiers. The next sections contain the
summary of each main program element  and the conclusions for each
of the frontiers.


\section{Intensity Frontier}
\label{sec:IF}

All frontiers of particle physics aim to discover and understand the
constituents of matter and their interactions at the highest energies, at the
shortest distances, and at the earliest times in the Universe. The Standard
Model (SM) fails to explain all observed phenomena: New interactions
and yet unseen
 particles must exist. They may manifest themselves either directly,
 as new particles, 
or by causing reaction rates to differ from the often very precise
predictions of the SM. The Intensity 
Frontier explores these fundamental questions by searching for new
physics in
extremely rare  processes or those forbidden in the SM. This requires the
greatest possible beam
 intensities, as well as massive, ultra-sensitive detectors. Many of
 these experiments
 are sensitive to new physics at higher mass scales or weaker
 interaction strengths
 than those directly accessible at high-energy colliders, thus
 providing opportunities
 for paradigm-changing new discoveries complementary to Energy and
 Cosmic 
Frontier experiments.

The range of experiments encompassing the Intensity Frontier is broad and
diverse.  Intense beams of neutrinos aimed over long distances at very large
detectors will allow us to explore the neutrino mass hierarchy, and 
search for CP violation and
non-standard interactions.  The very large detectors, if located
underground,
 will provide increased sensitivity to proton decay.
Multi-ton detectors searching for neutrinoless double-beta decay 
will determine whether neutrinos are their own
antiparticles. Intense beams of electrons will enable searches for
hidden-sector particles that may mediate dark matter interactions.
Extremely rare muon and tau decays, if seen, will signal violation of charged
lepton quantum numbers.
Measurements of intrinsic lepton properties, such as electric and magnetic
dipole moments, are another promising thrust. Rare and CP-violating decays of
bottom, charm, and strange particles, measured with unprecedented precision,
will clarify the new physics underlying discoveries at the Large
Hadron Collider (LHC).
In any new physics scenario, Intensity Frontier experiments with sensitivities
to very high mass scales will be crucial for exploration.

At Snowmass, the Intensity Frontier program was defined in terms of six areas
that formed the basis of working groups,  with experiments that probe 
neutrinos, baryon number violation, charged leptons,
quark flavor physics, nucleons, nuclei, and atoms, and new light,
weakly-coupled particles. 

The working group reports provide a clear overview of the science
program within each 
area of the Intensity Frontier.  They present discovery opportunities for facilities 
that will be available this decade or will come online during the next
decade.  Here,
 we briefly summarize the findings from each working group. 

\noindent{\bf Neutrinos:} 
Decades of experimental and observational scrutiny have revealed less than a
handful of phenomena outside the SM. These include the dark energy and dark
matter puzzles, and the existence of non-zero neutrino masses.  Neutrino masses
represent one of the few experimental pointers towards a new underlying theory.     
While many experiments continue to look for other new phenomena and
deviations
 from SM predictions, it is clear that continued detailed
 study of the 
neutrino sector is of the utmost importance.    

Compared to the other fermions, the elusive neutrinos have been extremely
difficult to study in detail.  Despite the challenges, neutrino physics
has been tremendously successful over the past two decades.  From almost 
complete lack of knowledge about neutrino mass and mixing twenty years ago,
we now have a robust, simple, three-flavor paradigm describing most of the
data. 

 However, key questions in the three-flavor sector remain:  We do not
know the mass ordering or the value of neutrino masses, 
nor whether neutrinos violate CP symmetry, nor whether
the neutrino is its own antiparticle, and we have only just begun to test the
three-flavor paradigm. A precision neutrino oscillation program is required to
carry out such measurements.  Furthermore, some experiments have uncovered
intriguing anomalies that merit additional study, and could lead to the
discovery of states or interactions beyond the SM.  Advances in
detector technology and analytical techniques for the next generation of
neutrino experiments are well underway.  We have clear experimental paths
forward for building on our success, for precision testing of the three-flavor
paradigm, for the exploration of anomalies, and for the measurement of fundamental
neutrino properties and interactions.

The next decade promises significant experimental progress around the world.
In the search for neutrinoless double-beta decay, a number of experiments rely
on complementary isotopes and experimental techniques.  The next generation of
100-kg-class neutrinoless double-beta-decay search experiments should have
sensitivity to effective masses in the 100~meV range; beyond that, there are
opportunities for ton-class experiments that will reach sub-10 meV
effective mass sensitivity, pushing below the inverted hierarchy region.  The next 
generation of tritium-beta-decay experiments will directly probe neutrino masses a 
factor of 10 smaller than the best current bounds.  Innovative ideas may help to go 
beyond these sensitivities.  

The neutrino mass hierarchy can be unambigously resolved using accelerator
neutrino oscillation experiments with baselines around 1000~km (or longer) and
detector masses of order tens of kilotons.  Precision measurements of
atmospheric neutrino oscillations with megaton-scale underground
detectors
or detectors under ice  can
also resolve the mass hierarchy.  The discovery of a non-zero
$\theta_{13}$
mixing angle
enables long-baseline neutrino experiments to search for leptonic CP violation
in appearance experiments.  The search for CP violation in the neutrino sector
is a top priority for particle physics efforts worldwide, and  vigorous
planning for the next-generation large-scale neutrino oscillation experiment is
underway internationally.  Regardless of the experimental approach, high-power
proton beams (greater than 1~MW) coupled with massive detectors (of order 100
kiloton), are needed to study CP violation in neutrino oscillations. 
The U.S., with the Long-Baseline Neutrino Experiment (LBNE) and a
future 
multi-megawatt beam from Project
X at Fermilab, is uniquely positioned to lead an international campaign to measure CP violation
and push the limits of the three-flavor paradigm. 
An underground location for a far detector significantly enhances the physics
reach.  LBNE represents a vibrant global program with the U.S. as host.

Given the challenges associated
with precision measurements in the neutrino sector, complementary
baselines,
 sources, and detector techniques will be required to bring the
 picture into focus.
New accelerator technologies, such as
neutrino factories and cyclotron-based-sources, may eventually
take measurements to the next level. Smaller
experiments will also play a key role in addressing some of the remaining
anomalies and hints for physics beyond the three-neutrino paradigm,
and 
study neutrino--matter interactions in detail.
 
The diversity of physics topics that can be probed through the
neutrino sector
 is very significant, and the interplay between neutrino physics and
 other fields 
is rich.   Neutrinos can and will provide important information on
structure 
formation in the early Universe; Earth, Sun, and supernova physics;
nuclear 
properties;  and rare decays of charged leptons and hadrons.  
{\it The
neutrino
sector sits at the nexus of a worldwide effort that crosses the 
frontiers of particle physics.}

\noindent{\bf Baryon number violation:}
Within the SM, protons are stable, as baryon number is assumed to be conserved.  However, 
baryon number is not a fundamental symmetry of the SM and is not conserved in many of its 
extensions.  In particular, baryon number violation is an essential ingredient for the creation 
of the observed asymmetry of matter over anti-matter in the Universe.
Grand unified theories (GUTs) predict that the proton decays with a lifetime in
excess of $10^{30}$ years, with the decay being mediated at scales of order
$10^{16}$ GeV.  Two important decay channels in GUTs are $p\to e^+\pi^0$ and
$p \to \bar\nu K^+$; several other modes are also possible.  The current
limits on the proton lifetime in these two channels are roughly
$10^{34}$~years and
$6\times 10^{33}$~years, respectively, a factor of 5 to 10  below
predictions in certain well-motivated GUT models.

The search for proton decay is carried out in detectors containing 
enough protons, and placed underground to reduce backgrounds.  Large neutrino
oscillation detectors are ideal for this task, and proton decay is an important
piece of their physics portfolio.  The  largest existing underground neutrino
experiment is the 22.5-kiloton Super-Kamiokande water Cherenkov detector. Future underground
neutrino experiments, such as a 34-kiloton  liquid argon 
time-projection chamber (LBNE) or a 560-kiloton
water Cherenkov detector (Hyper-Kamiokande)
can measure lifetimes on the order of GUT expectations
with exposures of roughly 10 years. Typical exposure of these experiments
could reach a sensitivity of $\tau(p\to e^+ \pi^0) < 10^{35}$ years and
$\tau(p\to\bar\nu K^+) < 3\times 10^{34}$ years.  

Neutron-antineutron oscillations would violate baryon number by two
units.  They are 
 searched for with a beam of free neutrons, in which  a neutron would
 transform
 into an
antineutron that annihilates in a distant detector.  Such oscillations are expected
in theories where baryogenesis occurs near or below the electroweak scale.
A proposed experiment at
Project X at Fermilab, using free neutrons from a 1 MW spallation
target,
could discover this phenomenon or 
improve existing limits on the oscillation probability by four orders of magnitude.

\noindent{\bf Charged leptons:}
The charged lepton experimental program offers significant discovery
opportunities in this decade's experiments and in even more sensitive
experiments possible with future facilities such as Project X at Fermilab.
Extremely sensitive searches for rare decays of muons and tau leptons, together
with precision measurements of their properties, will elucidate the scale and
dynamics of flavor generation or limit the scale of flavor generation to well
above $10^4$ TeV.
Any indication of charged lepton flavor violation (CLFV) would be an
indisputable discovery of new physics. Precision measurments of lepton 
flavor-conserving (LFC) processes can be used to verify predictions of the SM
and 
look for signs of new physics.

The experimental program consists of a large and diverse set of opportunities and includes multi-purpose experiments that utilize the large tau production rates at high-luminosity $B$ factories, as well as highly optimized single-purpose experiments that explore muon transitions.

Significant advances in studying CLFV in the muon sector can be achieved this decade.  For the rare 
decay $\mu\to e\gamma$, the  MEG upgrade at the Paul Scherrer Institute (PSI)
can reach branching fractions up to 
$6\times 10^{-14}$.  The Mu3e collaboration at PSI plans to improve their
sensitivity to $\mu\to 3e$ by approximately four orders of magnitude.  

Observation of 
the direct conversion of a muon to an electron in the field of a
nucleus would provide a powerful window
to physics beyond the SM.  Current limits for
$\mu N\to e N$ conversion are at the level of $10^{-12}$ to $10^{-13}$ from experiments at PSI.
Later this decade, COMET at J-PARC plans to improve these bounds by two orders of 
magnitude. A separate proposal at J-PARC, DeeMe, would use a different technique to reach
a similar sensitivity.  Before the end of the decade, Mu2e at Fermilab, followed soon by COMET,
will begin operations and improve the existing search reach by four orders of magnitude.
This would reach sensitivity to
signals from supersymmetric grand unified models.  If no signal is observed, 
this would set constraints on CLFV physics at the scale of $10^4$ TeV.  Future experiments beyond 
these are being considered in conjunction with more intense muon beams that could be available
with new facilities at J-PARC or Project X at Fermilab.

The muon's magnetic moment is predicted very precisely in the SM. New physics
contributes via radiative corrections.  The present level of sensitivity was
obtained by the E821 experiment at Brookhaven National Laboratory, with a
difference between the measurement and the SM theoretical prediction of $3.6
\sigma$.  A new experiment, E989 (Muon $g-2$) at Fermilab, will re-use the E821 muon storage ring at
Fermilab with the same experimental technique. E989 is expected to increase the
statistics by a factor of 20 with a corresponding reduction of systematic  uncertainties,
resulting in an overall reduction in the experimental error by a factor
of roughly 4. An alternate approach at J-PARC using lower-energy muons is
expected to have the same precision as E989 at Fermilab but 
very different systematics.  A world-wide effort is underway to reduce the theoretical
uncertainty in the SM prediction with new data
from $e^+e^-$ machines and breakthroughs in lattice gauge theory.

New
physics effects usually scale as a function of the lepton mass, and hence
$\tau$ observables can be very sensitive to new contributions. 
Important observables in $\tau$ leptons are CLFV decays, CP violation,
the electric dipole moment, and the anomalous magnetic dipole moment.  The large
$\tau$ production rates possible at the future SuperKEKB facility in Japan
could achieve an order of magnitude improvement in CLFV branching fractions
over current results from BABAR and Belle.

The charged lepton sector has
significant potential to reveal more information on the fundamental principles
of nature, and the U.S. has the opportunity to play a leading role with
facilities planned for this decade.

\noindent{\bf Quark flavor physics:}
The study of strange, charm, and bottom quark systems has a long and rich
history in particle physics. Measurements of rare processes in the flavor sector have led to 
startling revelations and played a critical role in the development of the SM.  The constraints 
on physics beyond the SM from flavor physics considerations are powerful.  The current quark 
flavor data set is mostly in agreement with SM expectations with a
handful of 3 to 4$\sigma$
anomalies.
New corrections to the SM at the level of 
tens of percent are still allowed by the data.  
Contributions to flavor processes from many theories beyond 
the SM arise at this level, and thus more precise measurements may observe
new physics.  If new massive states are observed at the LHC, 
detailed measurements of the quark flavor sector will be necessary to determine 
the underlying theory and its flavor structure.  If such states are not discovered in high-energy 
collisions, then precision quark flavor experiments, with their ability to probe mass scales far 
beyond the reach of the LHC, provide the best opportunity to set the next energy scale to explore. 
Depending on the strength of new physics interactions, this program already indicates that the new 
physics scale is above 1 TeV and in some scenarios above $10^5$ TeV. Proposed 
experiments can probe even further.  Continued 
investigations of the quark flavor sector are thus strongly motivated.

A well-planned program of flavor physics experiments has the potential to continue this history 
of advances.  Such a program exists worldwide with the LHCb experiment at the LHC, an upgraded 
SuperKEKB facility in Japan, BESIII in China, and future rare kaon decay experiments at CERN, 
J-PARC and potentially Fermilab.  These facilities will carry out a rich 
multi-purpose program in the strange, charm, and bottom sectors and
perform
 numerous crucial measurements
of rare decays and CP-violating observables.  The proposed experiment ORKA at
the Fermilab Main Injector would probe rare kaon decays to unprecedented precision and would 
retain the U.S. capability to perform quark flavor experiments.  In
the longer term, Project X at 
Fermilab could become the dominant facility in the world for rare kaon decays. The expected 
sensitivities for these future programs are detailed in the full report, and are at the level 
which could discover new physics.  It is important to note that these results are not predicated 
on future theoretical progress, although theoretical advancements will strengthen the program by 
increasing the set of observables that can reveal new physics.  
U.S. contributions and support for quark flavor experiments are necessary in order for the U.S. 
HEP program to have the breadth to assure meaningful participation in future discoveries.

\noindent{\bf Nucleons, nuclei, and atoms:}
The use of nucleons, nuclei, and atoms as laboratories for the study of
fundamental interactions is entering a new era.  These systems have sensitivity
to physics beyond the SM and provide important tests of vital symmetries
through measurements of electric dipole moments (EDMs), weak decays of light
hadrons, weak neutral currents, and atomic parity violation.  

Observation of an EDM would signify both parity and
time-reversal symmetry violation, and would probe  the physics of CP violation.
The SM predictions (via multi-loop contributions) for the EDMs of the electron,
neutron, and nucleus are $10^{-38}$, $10^{-31}$, and $10^{-33}$
{\it e}$\cdot$cm, respectively.  EDM measurements are challenging
and the present experimental sensitivity is approximately $10^{-27}$,
$3\times 10^{-26}$, and $3\times 10^{-29}$ {\it e}$\cdot$cm for the electron,
nucleon, and $^{199}$Hg nucleus, respectively. Experiments searching for the
electron EDM typically use the polar molecules YbF and ThO and ultimately
expect to reach a level of $3 \times 10^{-31}$ {\it e}$\cdot$cm.  Several
current or planned experiments searching for the neutron EDM are expected
to reach a sensitivity of $5\times 10^{-28}$ {\it e}$\cdot$cm, a
factor of 100 below current limits.  For atoms, future experiments
using mercury, radon, and radium expect sensitivities at the level of $10^{-32}$
{\it e}$\cdot$cm.  This would require upgraded facilities such as FRIB at
Michigan State University or Project X at Fermilab.  These future programs will
be sensitive to signals predicted to appear in several theories beyond the SM.

The weak decays of light hadrons provide precision input to the SM and are a
sensitive test of new interactions. The ratio of decay channels $e\nu/\mu\nu$
for pions and kaons affords a precise test of lepton universality, probing
new physics up to $10^3$ TeV.  For pions, experiments at TRIUMF and PSI
will improve the measurement error by a factor of five.
Neutron beta decay provides an accurate determination of
the CKM element $V_{ud}$, enabling strong tests of CKM unitarity and constraining new physics up to scales of $\sim 10$ TeV.  Several programs are underway to measure observables of the neutron lifetime and decay asymmetries with improved precision.
Measurements of parity-violating asymmetries in fixed-target scattering with polarized electrons allows for a precision determination of the weak mixing angle at low values of momentum transfer, which in turn constrains new parity-violating effects up to $2-3$ TeV.  An improved polarized M{\o}ller scattering experiment
at upgraded JLab facilities expects to determine the weak mixing angle to an
accuracy of 0.1\%,
comparable to measurements at the $Z$ pole.  Parity violation in atomic
transitions also yields valuable measurements of the weak mixing angle.  New
techniques requiring intense sources are being developed; Project X at Fermilab
would provide more rare isotopes for this program than any other facility.

\noindent{\bf New light, weakly coupled particles:}
New light particles that couple very weakly to the SM fields are a common
feature of extensions beyond the SM.  They are motivated by both theoretical
and observational considerations, including the strong CP problem and the
nature of dark matter and dark energy.  Examples of such particles include
axions, hidden-sector photons, milli-charged particles, and chameleons.  These
hidden-sector particles typically couple weakly to the photon via mixing.
Intense sources are hence
required to produce them at rates sufficient to enable their discovery.
The parameters relevant for searches of such hidden-sector particles are their
mass and coupling strength to the photon.  A variety of experiments constrain
part of this parameter space, but much territory is still open for
exploration.
 The current constraints arise from astronomical
observations, cosmological arguments, and a variety of 
laser, heavy-flavor, and fixed-target experiments.
Regions that may signal dark matter detection or annihilation and areas that offer
an explanation for the present result on the anomalous magnetic moment of the
muon have yet to be probed.

Numerous laboratory
experiments are either in progress or proposed.  Two microwave cavity
searches for axions 
will be underway soon in the U.S., but require further developments
to increase their mass 
reach.  The light-shining-through-walls technique, where photons are
injected against an
 opaque barrier, continues to explore open regions of parameter space.
 More advanced
 technology is needed to make progress in the mid-term.  Axion
 helioscope searches 
were first carried out using borrowed magnets, and now require a
custom-built magnet
 to improve sensitivity.  Collider searches for hidden-sector
 particles can 
be performed via the reaction $e^+e^-\to\gamma\ell^+\ell^-$ at high 
luminosity $e^+e^-$ factories or in the decays of gauge bosons at the
LHC.  
Fixed-target experiments using both electron and proton beams are a
promising
 place to search for hidden-sector particles.  The electron beam
 experiments APEX at the  Jefferson National Accelerator Facility
 (JLab) and A1 at the University of
 Mainz have recently performed short test runs and have plans for more
 extensive runs this decade.  HPS has been approved by JLab and will
 run after the 12 GeV upgrade.  DarkLight
 proposes to use the free-electron laser beam at JLab.  Proton
 fixed-target experiments have the potential to explore regions of
 parameter space that cannot be probed by any other technique. 
 Neutrino experiments, such as MiniBooNe, can widen the search to
 smaller couplings.
The intense proton source at Project X could also provide a 
powerful extension to the search reach.   
Impressively large regions of parameter space are currently unexplored
and are
 ripe for the discovery of light, weakly-coupled particles.

\noindent{\bf Conclusions:}
The above program exhibits the broad spectrum of science opportunities
attainable at the Intensity Frontier.  While each subfield is at a
different stage of maturity in terms of testing the SM, the proposed
experiments in each area are poised to have major impact.  The
programs involving transitions of heavy quarks, charged leptons, and
nucleons, nuclei, and atoms are advanced, with the most precise SM
predictions and a well-developed experimental effort that has spanned
decades.  In this case, the next level of experimental precision would
reach the point where effects of new TeV-scale interactions are
expected to be observable.  More sensitive searches for proton decay
and new light, weakly-coupled particles can cover a large range of
parameter space that is consistent with grand unified theories and
cosmological observations.  Neutrino physics is just beginning the era
of precision measurements where it is possible to probe basic neutrino
properties and answer principal questions.  Neutrino physics holds
great promise for
 discovery. 

Such an extensive program is necessary to address the unresolved fundamental
questions about nature.  The knowledge we seek cannot be gained by a
single experiment or on a single frontier, but rather from the
combination of results from many distinct approaches working together
in concert.  The full report from the Intensity Frontier 
provides a reference for the captivating
science that can be carried out in this decade
and next.  The Intensity Frontier program has the potential to make
 discoveries that change paradigms and alter our view of the Universe.


\def\MSbar{$\overline{MS}$}
\def\sstw{$\sin^2 \theta_W$}
\section{Energy Frontier}
\label{sec:energy}

Experiments at the Energy Frontier make use of high-energy accelerators to 
produce and study heavy elementary particles and to search for new ones.   The
Energy Frontier includes experiments at the Large Hadron Collider at CERN
and those at future colliding-beam accelerators proposed for lepton-lepton and
proton-proton collisions.

The first run of the LHC has closed a nearly half-century-old chapter in the
story of elementary particle physics. We have discovered a most unusual new
particle with properties very similar to those expected of the Standard Model
Higgs boson.  The appearance of this particle --- and further confirmation of
its identity --- ends one era and opens another. On one hand, the Standard Model
of particle physics is complete. We know all of the particles in this model and
how they interact with one another and we have at least a basic idea of their
properties.  On the other hand, we also know that the Standard Model is 
incomplete in important ways.  It challenges us to uncover the physics behind
its apparently ad hoc structure. We are certain that a host of observed
anomalous phenomena and set of confusing conceptual questions have explanations
that require new physics outside the Standard Model.

The LHC and the CMS, ATLAS, and LHCb detectors have brought to bear impressive
capabilities for exploring  the answers to these new questions.  The LHC
accelerator is expected to dramatically increase its ability to deliver beams
in the period between now and 2030, increasing its energy by almost a factor of
two and its integrated luminosity by a factor of 100.  The detectors will
improve their ability to collect enormous data sets and to discriminate the
properties of events with increasing precision. Around the world, other new
accelerators are being considered that will give us additional power in
understanding the heaviest particles of the Standard Model and exploring for
new ones.  In this report, and in the detailed working group reports, we trace
out the programs of these accelerators and present their most important goals. 

{\bf Importance of the TeV energy scale:}
Our successful theory of weak interactions is based on the idea of an
underlying symmetry that is spontaneously broken.  The symmetry of the theory
of weak interactions dictates the couplings of the quarks and leptons to the
$W$ and $Z$ bosons.  Its predictions have been confirmed by high-precision
experiments. However, this symmetry forbids the quarks, leptons, and vector
bosons from having mass. To reconcile the symmetry of weak interactions with
the reality of particle masses, one more unexpected element is required.  This
is a field or set of fields that couples to all types of particles and forms a
condensate filling the Universe.  The discovery of the Higgs particle
establishes that this condensate exists and is the origin of particle masses.

This is an historic achievement. It is not an end but a beginning.  It
highlights many questions that the Standard Model leaves unanswered.  These
require new, equally bold ideas.  Two of these questions --- the nature
of the Higgs field and the composition of dark matter --- give particularly
strong motivations for collider experiments.

The Standard Model does not explain the underlying structure of the Higgs field
or the reason why it condenses. It does not explain the magnitude of the
condensate, which sets the mass scale of all known elementary particles.  The
fact that the observed Higgs particle is a scalar particle makes it very
difficult to understand why this scale is smaller than other basic mass scales
of nature such as the Planck scale.  There are no simple models that answer
this question.  New fundamental structures are needed.  The Higgs field must be
a composite of more basic entities, or space-time itself must be extended,
through supersymmetry or through extra dimensions of space.  These ideas
predict a rich spectrum of new elementary particles, typically including a
larger set of Higgs bosons,  with masses at the TeV energy scale.

The Standard Model also does not account for the dark matter that makes up most of
the matter of the Universe.  The simplest and most compelling model of dark
matter is that it is composed of a stable, weakly interacting, massive particle
(WIMP) that was produced in the hot early Universe.  To obtain the observed
density of dark matter, this model requires the WIMP interactions to be roughly
at the TeV energy scale.  If this model is correct, it may be possible to study
dark matter under controlled laboratory conditions in collider experiments.

\emph{Compelling ideas about fundamental physics predict new particles at
the TeV energy scale that should be discoverable in experiments at the LHC and
planned future accelerators.  These experiments will provide the crucial tests
of those ideas. Furthermore, if such particles are discovered, they can be
studied in detail in collider experiments to determine their properties and to
establish new fundamental laws of nature.}

The past successes of particle physics and its current central questions then
call for a three-pronged program of research in collider experiments:
\begin{enumerate}
\item We must study the Higgs boson itself in as much detail as possible,
searching for signs of a larger Higgs sector and the effects of new heavy
particles.
\item We must search for the imprint of the Higgs boson and its possible
partners on the couplings of the $W$ and $Z$ bosons and the top quark.
\item We must search directly for new particles with TeV masses that can
address important problems in fundamental physics.
\end{enumerate}
The Energy Frontier study pointed to all three of these approaches as
motivations for further experiments at colliders. The results of the study
confirmed that the existing LHC detectors and their planned upgrades, together
with proposed precision lepton collider experiments, will be nimble and
sensitive enough to carry this three-fold campaign forward into the next two
decades.

The Energy Frontier study was organized into six working groups --- on
the Higgs boson, the $W$ and $Z$ bosons, quantum chromodynamics (QCD),
the top quark, new particles and forces, and flavor interactions at high energies.
Each working group was asked to evaluate the
future program for its topic both from a high-level perspective and from the
viewpoint of supplying motivation for experiments at a range of proposed
accelerators. In the remainder of this section, we present the conclusions
of these reports, first by physics topic, then by facility.

{\bf Higgs boson:}
A new bosonic resonance at 125~GeV was discovered
at the LHC only one year ago.  Many properties of this particle have now been
measured  and, up to this point, are consistent with those of the Higgs boson
of the minimal SM. The couplings of this boson roughly scale with
mass.  The specific form of the coupling to the $Z$ boson indicates that the
particle has spin-parity $0^+$ and that the corresponding field has a nonzero
vacuum expectation value.

However, we cannot be complacent about the identity and role of this particle.
On one hand, the idea that a single scalar field is solely responsible for the
generation of all particle masses is just one possibility among many and needs
explicit verification.  On the other hand, models with additional Higgs bosons
and related new particles, and models in which the Higgs boson is composite,
are hardly tested.  Deviations from the minimal Higgs boson properties due to
new particles with mass $M$ are suppressed by a factor $(m_h/M)^2$, so to the
extent that the LHC has set lower limits on the masses of new particles at many
hundreds of GeV, we would not yet have expected to see the modifications to the
Higgs properties caused by those particles.  

An experimental program to probe the Higgs boson contains several elements.
The first is to search for deviations from the minimal SM
expectation that the Higgs boson couples to each particle species according to
its mass.  Such deviations are expected in almost all models of new physics.
However, the effects are expected to be small, at the few-percent level if
induced by new particles that will not be directly detected at the LHC.  There
is a characteristic pattern of deviations for each new physics model.  The
High-Luminosity LHC (HL-LHC) is expected to measure these couplings with
precisions of several percent, varying from coupling to coupling.  Lepton
collider experiments have the potential to push these precisions to the
sub-percent level, which would be needed to uncover deviations from Standard
Model predictions with significance high enough to claim evidence of new
physics.

Such  a program of precision measurements of Higgs couplings requires a parallel
concerted effort in precision theory.  It also requires improvement of our
knowledge of crucial input parameters such as $\alpha_s$ and $m_b$, which can
be provided by lattice gauge theory computations.  Collider experiments can
also probe the nonlinear Higgs field self-coupling to the 10--20\% level,
thereby testing the critically important question of the shape of the Higgs
potential.

Future experiments should also improve our knowledge of the Higgs
boson 
mass and quantum numbers.  The spin of the observed resonance should
already be clear from LHC data in this decade.  A  more subtle question is
whether this particle contains a small admixture of a CP-odd state, signaling
CP violation in the Higgs sector and confirmation of at least one additional
Higgs-like particle.  We discuss probes for this effect at various colliders.

Finally, it is important to search directly for additional Higgs bosons.  The
LHC can probe to masses of 1 TeV with model-dependent limits.  Lepton colliders
can can make more model-independent searches to masses close to the collider beam
energy.

{\bf $W$ and $Z$ boson, QCD, and the top quark:}
The study of $W$ and $Z$ bosons includes both the extension of the
program of precision electroweak measurements, and the search for
new interactions in the  three- and four-vector boson couplings.

The minimal SM makes precise predictions for the well-studied
precision observables $M_W$ and \sstw.  At the moment, the observed values
are within 2$\sigma$ of the predictions; the deviations are consistent with
the effects of new particles in a range of new physics models.  Better
precision in this program is clearly needed.  Future experiments will sharpen
our knowledge of these quantities and potentially expose inconsistency with the
SM.  The LHC, especially in its high-luminosity phase, has the
potential to reduce the error on the $W$ mass to $\pm$5~MeV.  This requires a
factor of 7 decrease in the current error due to parton distribution functions
and is a challenge to QCD researchers.  Lepton colliders can make further
improvements, to an error of $\pm$2.5~MeV, with a dedicated measurement of the
$WW$ threshold.  A linear collider with beam polarization running at the $Z$
resonance to produce $10^9$ $Z$ bosons (Giga-Z) is expected to reduce the error
on \sstw\ by a factor of 10.  Finally, a circular $e^+e^-$collider operating
in a 100 km tunnel can potentially push both errors down by another factor of
4. All of these precision measurements challenge the
inflexible correlations among the SM particles and their respective
forces.  Such precision measurements of  electroweak observables could
become  discoveries of new physics if 
 the tight constraints within the SM begin to
unravel.

The second theme of $W$ and $Z$ boson studies is the search for anomalous nonlinear
couplings of the vector bosons.  Collider experiments with enough
energy to produce pairs of 
$W$ and $Z$ bosons are 
sensitive to three-gauge-boson couplings.  At the LHC,  we will be 
sensitive, for the first time, to
non-standard four-boson interactions, which would indicate
 new interactions in vector
boson scattering.  Lepton collider experiments have the potential to push
current uncertainties on three-boson couplings down by an order of magnitude,
into the region in which new physics effects are predicted in models in which
the Higgs boson is composite.   Both hadron and lepton colliders can access
vector boson scattering, but the total center-of-mass energy available in a
scattering process is a crucial factor. The high-luminosity LHC will be sensitive to
vector boson or Higgs resonances with masses well above 1 TeV.

QCD is well established as the correct theory of the
strong interactions.  Nevertheless, advances in QCD are needed to achieve the
goals of future experiments, especially at hadron colliders.  These
experiments
require improved knowledge of the parton distribution functions.
That  can be achieved with data expected from the LHC on the rapidity
distributions of $W$, $Z$, and top quark production.  In addition, precision
cross-section computations, to the NNLO level, are needed for many two- and
three-particle production processes, especially those involving the Higgs boson.
This will require advances in the theoretical art of QCD computation.  Finally,
it is important to push the error on the value of $\alpha_s$ below the 0.5\%
level.  Lattice gauge theory seems to be a promising avenue for
achieving this goal.

The top quark was discovered at the Fermilab Tevatron and studied there with
samples of tens of thousands of $t\bar t$ pairs.  The LHC experiments
will
produce and 
study
billions of  top
quarks.  At future lepton colliders, we will use the
electroweak couplings of top quarks as a production mode and probe these with
polarization observables.  Both methods will transform our knowledge of this
quark, whose properties are intimately connected to the mysteries of flavor and
mass generation.  To this day, we are surprised at the high mass of this
presumably fundamental particle and its proximity to the value of the Higgs
vacuum expectation value. 

The top quark mass is not only an important puzzle in itself but also is an
important input parameter for particle physics.  The strongest demands on
precision in the top quark mass come from the precision electroweak program,
where interpretation of a 5 MeV error in $m_W$ requires a 500~MeV error on
$m_t$.  This mass must be a theoretically well-defined quantity, convertible to
a short-distance parameter such as the \MSbar\  mass.  There are strategies
applicable at the LHC that allow the measurement of a well-defined top quark
mass to this 500~MeV accuracy.  At lepton colliders, measurement of the cross
section at the top quark pair production threshold gives the \MSbar\ mass to
100~MeV, as required for the more accurate precision electroweak program
available at these machines.

Top-quark couplings will be studied with high accuracy both at hadron and at
lepton colliders.  New physics from top quark and Higgs compositeness can
create few-percent corrections to the gluon, photon, and, especially, $Z$ boson
couplings.  These effects can be observed as corrections to the pair-production
cross sections relative to  the predictions of the SM.  The
top-quark coupling capabilities of a lepton collider are especially strong,
with accuracies possible at the sub-percent level.  The billions of top quarks
produced at the high-luminosity LHC allow very sensitive studies of rare
flavor-changing top decays, to a level that complements searches at low energy
for flavor-changing quark decays. 

Models of the Higgs potential and its symmetry breaking typically require new
particles that are partners, in some way, of the top quark. The LHC, especially
in its high-luminosity stage, will have the capability for extensive searches for
supersymmetric partners of the top quark, heavy vector-like top quarks that
appear in models with Higgs and top quark compositeness, and heavy resonances
that decay to $t\bar t$, which appear in models with new space dimensions.

{\bf Searches for new particles and interactions:}
High-energy colliders can search for new particles with a very
broad range of properties.   These particles,  with masses near the 1
TeV scale,
are required in models of electroweak symmetry
breaking.  Other questions also call for new particles accessible to high
energy colliders.  A large class of models of dark matter place the dark matter
particle as the lightest particle of a TeV mass spectroscopy.  Grand
unification requires new particles near the TeV scale, including partners of
known particles and perhaps also new vector bosons associated with enhanced
gauge symmetry.  CP violation in the Higgs boson sector is required in models
that generate the matter-antimatter asymmetry at the electroweak phase
transition.  More generally, new particles can bring new sources of flavor and
CP violation that might be reflected in the discovery of new flavor-changing
reactions at low energy.

The LHC has already, in only its first run, increased the reach and power of
searches for new particles over a broad scope.   We expect that this power will
increase dramatically in the next decade, as the LHC experiments acquire 300
fb$^{-1}$ of data at 14 TeV.  This extension probes deeply into the region
expected for the masses of new particles in all classes of models of electroweak
symmetry breaking.  The
high-luminosity stage of the LHC, up to 3000 fb$^{-1}$, will provide a further,
very significant, extension of the search region.  This extension is
particularly powerful for states produced through electroweak interactions, for
which a factor of 2 increase in the mass reach is available in some
cases.

Lepton colliders would bring new and complementary capabilities.  They would
carry out model-independent searches for states such as dark matter candidate
particles whose signatures are especially difficult to observe at hadron colliders.
Lepton colliders
 would uncover new decay modes and measure branching ratios and quantum
numbers for any new particle within their energy range.

{\bf Physics opportunities for colliders:}
The physics opportunities described above are reflected as motivations
for current and future high-energy colliders.  Our study considered a wide
range of proposed machines. The full report from the Energy Frontier presents
the cases for these machines in some detail. 

We first point out the opportunity provided by the 14 TeV run of the LHC
scheduled for the next decade.  This will provide robust searches for new
particles over a broad front, with great promise of the discovery of the TeV
particle spectrum motivated at the beginning of this section. Any plan
for high energy physics in the longer
term must include the possibility of discovering new particles
 in this period and exploiting that discovery at the facilities that will follow. 

We find the case for the high-luminosity stage of the LHC compelling. This plan
to deliver 3000 fb$^{-1}$\ has been listed in the European Strategy
for Particle Physics as
the highest priority accelerator project in Europe for the 2020's.  We find
that it will provide a significant additional step in the search for new
particles, and that it will provide other important capabilities.  The most
important of these is the beginning of the era of precision Higgs boson
measurements, to few-percent precision.  It is likely to give the first
evidence of the Higgs boson self-coupling.  It will provide a program of
precision measurement in the SM that will dramatically tighten our
knowledge of the $W$ boson and the top quark, with measurements
sensitive
to the predictions of a variety
of new physics models.  We have already noted that the additional luminosity
will significantly enhance the capability of the LHC to search for new heavy
particles.

We considered the scientific case for the International Linear Collider (ILC).
This next-stage lepton collider has recently completed its Technical Design
Report and was judged in the Snowmass study to be ready for construction.  This facility
is named as the highest priority for new initiatives by the Japanese
particle
physics community.  We find that this machine is strongly
motivated.  It will reach sub-percent accuracy in the study of the Higgs boson,
allowing discovery of percent-level effects in the Higgs couplings predicted in
new physics models.  It will measure the Higgs width in a model-independent
way.  It will give the capability to observe all possible Higgs modes,
including decays to SM modes not observable at the LHC, to dark
matter, and to other invisible and exotic states.  It will extend our knowledge
of the top quark and the $W$ and $Z$ bosons well beyond the precision achievable at
the LHC, setting up a confrontation with models that include Higgs boson and
top quark composite structure. 

The Energy Frontier study considered many other accelerator facilities for
construction over longer time scales.  These included higher-energy linear
colliders, circular $\ee$ colliders, muon colliders, and photon colliders.  We
present a detailed discussion of the physics motivations for these facilities
in our full report.  There was particular interest in a proton collider of
energy 100 TeV (VLHC), which would come close to the capability of covering the
full model space for models of ``natural" electroweak symmetry breaking and
WIMP dark matter.  Our study developed materials and resources to begin a more
complete survey of physics at such a high-energy collider.  This study, and a
parallel development of magnet technology for higher-energy proton colliders,
should be pursued over the next decade.

{\bf Conclusions:}
Previous surveys of the prospects for high-energy accelerator experiments have
spoken in terms of reducing the space of parameters --- couplings, mixings,
masses --- as if that were the goal.  Now, more than ever, the momentum points
not toward exclusion, but toward the discovery of new states.  Many possible
directions are open and must be pursued.  

The Higgs boson discovery changes everything.  It transforms the research agenda
for particle physics, giving us a set of sharp questions that we
cannot ignore.
It motivates more strongly the exploration of the TeV energy
scale, where the solution to the mystery of dark matter and other key
problems might also be found.
The study of the Higgs  particle to high precision, together with
high-precision studies of the $W$, $Z$, and top quark and searches for new
states, provide us with complementary routes to fully explore the 
particles and forces in this range of energies.
 The current LHC detectors and their planned upgrades are
well suited to carry on this program.  Future accelerators will bring new
capabilities to pursue it further.

High-energy colliders provide manifest opportunities to discover new
fundamental interactions of broad consequence.  U.S. physicists have
been
leaders in Energy Frontier experiments up to now and are well
positioned to take a
leading role in the discoveries of the coming decades.
\section{Cosmic Frontier}
\label{sec:Cosmic}

Investigations at the Cosmic Frontier use the Universe as a laboratory
to learn about particle physics.  Our understanding of the Universe
has been transformed in recent years.  In particular, experiments at
the Cosmic Frontier have demonstrated that only 5\% of the contents of
the Universe are well understood, with the rest composed of mysterious
dark matter and dark energy.  As a result, the Cosmic Frontier now
plays a central role in the global particle physics program, providing
overwhelming evidence for new particles and new interactions, as well
as powerful, unique opportunities to address many of our most
fascinating questions: What is dark matter?  What is dark energy?  Why
is there more matter than antimatter?  What are the properties of
neutrinos?  How did the Universe begin?  What is the physics of the
Universe at the highest energies?

To identify outstanding scientific opportunities for the coming 10 to
20 years, the Cosmic Frontier Working Group was organized into six
subgroups: Weakly-interacting massive particle (WIMP) 
 dark matter direct detection, WIMP dark matter
indirect detection, non-WIMP dark matter, dark matter complementarity,
dark energy and CMB, and cosmic particles and fundamental physics.  In
several cases, these subgroups were further divided into topical
working groups. 

The $\Lambda$CDM standard model of cosmology provides the backdrop for
much of Cosmic Frontier research.  In this model, the Universe
underwent a very early epoch of accelerated expansion (inflation),
which was followed by eras in which the Universe was dominated
successively by radiation, cold dark matter (CDM), and dark energy
($\Lambda$).  At present, the known particles make up only 5\% of the
energy density of the Universe, with neutrinos contributing at least 0.1\%.
The rest is 25\% dark matter and
70\% dark energy.  Remarkably, incisive measurements that explore all
of the key components of the model are now within reach.  The leaps in
sensitivity of the new facilities bring us to a time with strong
discovery potential in many areas.  Further surprises are likely in
this rapidly advancing area, with potentially far-reaching
consequences.

 {\bf Dark matter:}
The work of Snowmass highlighted the coming decade as one of
particular promise for the goal of identifying dark matter.  Evidence
for particle dark matter has been building for 80 years through the
study of galaxy clusters, galactic rotation curves, weak lensing,
strong lensing, hot gas in galaxy clusters, galaxy cluster collisions,
supernovae, and the cosmic microwave background (CMB).  However, all
evidence so far is based on dark matter's gravitational interactions,
and its particle identity remains a deep mystery.

Among the many dark matter candidates, one well-known possibility is
weakly-interacting massive particles with masses in the 1 GeV
to 100 TeV range.  Particles with these properties
appear in many models designed to
address the gauge hierarchy problem.  In cosmology, particles with 
these properties may
obtain the correct relic density either through thermal freeze-out or
through an asymmetry connecting their number density to that of
baryons.

WIMP direct detection experiments search for the interactions of WIMPs
with normal matter.  WIMPs may scatter elastically off nuclei,
producing recoil energies in the 1--100 keV range, which can be
detected through phonons, ionization, scintillation, or other methods.
There are daunting backgrounds, and direct detection experiments must
be placed deep underground.  In the last several years, however, this
field has seen a burgeoning of innovative approaches to discriminate
signal from background, including experiments incorporating dual-phase
media, self-shielding, pulse shape discrimination, and threshold
detectors.

The first two decades of direct detection experiments have yielded a
diverse and successful program, resulting in ``Moore's Law''-type
progress, with sensitivities doubling roughly every 18 months.  In the
coming decade, this rate of progress is expected to continue or even
accelerate for both spin-independent and spin-dependent interactions.
Upcoming second-generation (G2) experiments will improve sensitivities
by an order of magnitude, probing the Higgs-mediated cross sections
expected for well-known supersymmetric and extra-dimensional
candidates, and also extending the sensitivity to both $\sim \gev$
low-mass WIMPs, where possible signals have been reported, and $\sim
\tev$ masses that are beyond the reach of colliders. Following these
experiments, multi-ton-scale third-generation (G3) experiments are
expected to improve current sensitivities by up to three orders of
magnitude and will either find dark matter or detect background events
from solar, atmospheric, and diffuse supernovae neutrinos.  Probing
beyond this sensitivity will require either background subtraction or
techniques such as directional detection or annual modulation.  The
Snowmass process produced a detailed census of present and proposed
direct detection facilities, with uniform treatment of their
capabilities and issues, along with a survey of promising
technologies.

WIMPs may also be found through indirect detection, in which pairs of
WIMPs annihilate, producing SM particles, including gamma
rays, neutrinos, electrons and positrons, protons and antiprotons, and
deuterons and antideuterons.  Detection of these particles may be used
to constrain or infer dark matter properties.  The expectation that
WIMP annihilation in the early Universe determines the dark matter
abundance sets a natural velocity-averaged annihilation cross section
of $\langle \sigma_{\rm{an}} v \rangle \sim
 3 \times 10^{-26}$ cm$^3$s$^{-1}$ for indirect detection experiments.

Gamma rays from dark matter annihilation may be detected by both
space- and ground-based experiments.  In space, the Fermi-LAT has
recently demonstrated the promise of this approach, excluding the
natural cross section $\langle \sigma_{\rm{an}} v \rangle$ for dark
matter masses below 30 GeV, given certain halo profile and
annihilation channel assumptions. The reach is expected to be
extended significantly with additional data.  On the ground, VERITAS
and other atmospheric Cherenkov telescopes have set significant limits
on dark matter properties
by looking for gamma rays from dark matter-rich dwarf galaxies.
Moving forward, the atmospheric Cherenkov telescope community has
coalesced to build the Cherenkov Telescope Array (CTA), with
sensitivity at the natural cross-section scale for dark matter masses
from 100 GeV to 10 TeV, far beyond current or planned colliders.
  These projections require U.S. involvement in CTA, which
will double the planned mid-sized telescope array and enable critical
improvements in sensitivity and angular resolution.

Neutrinos also provide promising means for indirect detection of dark
matter.  High-energy neutrinos from the core of the Sun would be a
smoking-gun signal of dark matter particle annihilation.  The signal
depends primarily on the spin-dependent WIMP-nucleon scattering cross
section, which determines the capture rate.  Current bounds from
Super-K in Japan and IceCube at the South Pole
 already provide leading limits on this cross
section, and PINGU, an infill array upgrade to IceCube, will extend
the sensitivity to lower masses.  In the coming decade, IceCube and
PINGU, along with Hyper-Kamiokande, will probe cross sections one to
two orders of magnitude below current bounds, with sensitivities
competitive with those of planned G2 direct detection experiments.

Antimatter signals of dark matter are pursued in a variety of ways.
Recent measurements of cosmic-ray positrons by the
AMS-02 magnetic spectrometer confirm and improve with excellent
precision earlier measurements by PAMELA and Fermi.  The rising
positron fraction could be indicative of positrons created in the
decay or annihilation of dark matter.  In the near future, AMS-02 will
extend its determination of the positron fraction to energies close to
1 TeV, and add important information on cosmic-ray propagation.  Given
the possibility of astrophysical sources of primary positrons,
however, it may be very difficult to definitively attribute the excess
positrons to dark matter.  Antideuterons provide a signal that is
potentially more easily discriminated from astrophysical
backgrounds. With a long-duration balloon flight, the General
Antiparticle Spectrometer (GAPS) detector could provide
sensitivities comparable to AMS-02.  Last, the production of positrons
and electrons from dark matter annihilation also produces secondary
radiation.  Detection of signals with radio to X-ray frequencies has
the potential to probe the WIMP parameter space.

The Snowmass process also evaluated the prospects for non-WIMP
candidates, which could be some or all of the dark matter.  The axion
is particularly well-motivated, as it arises from the leading solution
to the strong CP problem of the SM.  RF-cavity and solar
searches for axions, such as ADMX and IAXO, will probe a large range
of axion parameter space, including the cosmologically favored region,
and have strong discovery potential.  Sterile neutrinos are also
highly motivated by the observed non-zero masses of active neutrinos.
In the mass range where sterile neutrinos are dark matter candidates,
their radiative decays produce a monoenergetic photon, which may be
detected with X-ray telescopes.  Many other dark matter candidates
were also surveyed, including asymmetric dark matter, primordial black
holes, Q-balls, self-interacting dark matter, superheavy dark matter,
and superWIMP dark matter.

How do the diverse strategies for identifying dark matter fit
together? The Snowmass process produced a clear articulation of how
the different approaches --- including the direct and indirect detection
experiments mentioned above, but also particle colliders and
astrophysical probes --- each provide unique and necessary information.
This complementarity was examined in two theoretical frameworks.
First, the discovery prospects were examined in complete
supersymmetric models, with randomly selected parameters in the
phenomenological MSSM framework.  Second, the possibility that only
the dark matter particle is kinematically accessible was considered
using the framework of dark matter effective theories.  In both cases,
the complementarity of different approaches was evident at all levels,
both to establish a compelling dark matter signal and, just as
importantly, after discovery, to determine the detailed properties of
the particle or particles that make up dark matter.

 {\bf Dark energy and CMB:}
Cosmic surveys --- optical imaging and spectroscopic surveys and
detailed measurements of the CMB --- precisely map the Universe on many
different angular scales and over wide ranges of cosmic time.  They
provide unique information about cosmology and new physics, including
inflation, dark matter, dark energy, and neutrino properties.  These
measurements are challenging, requiring advances in instrumentation
and excellent control of systematic effects.  Fortunately, these
advances are now within reach, thanks to decades of investment and
close collaborations between particle physicists and astrophysicists.
The payoffs for this effort are large.

Measurements of the distance-redshift relation, first using supernovae
and then additional complementary techniques, revealed the expansion
history of the Universe, particularly over the past several billion
years, and yielded the surprising discovery that the expansion rate
has been increasing instead of decreasing.  Now we must determine what
is causing the cosmic acceleration.  This ``dark energy'' must produce
negative pressure to be responsible for the observed effect.  One
important clue is whether the negative pressure has been constant in
time or is evolving.  The stage III (the DES and HSC imaging surveys,
and the PFS and eBOSS spectroscopic surveys) and stage IV (LSST
imaging survey and DESI spectroscopic survey on mountaintops; Euclid
and WFIRST-AFTA in space) dark energy facilities will constrain both
the value and the evolution of the value with much higher precision,
as recommended in previous community studies, but they will also do
much more.  We must also check whether our description of gravity is
correct, and this is where measurements of the growth of structure,
over a wide range of distance scales using both imaging and
spectroscopic surveys, are needed.

There are several alternatives to general relativity (GR) that can
accurately describe the observed distance-redshift relation, but they
also modify the behavior of gravity over different distance scales.
The alternative models therefore predict structure growth rates that
are different from those in the standard theory.  Measuring the
structure growth rate over many different distance scales will test GR
and the alternative models.  Deviation from expectation on just one of
these scales will signal new physics.  In other words, the upcoming
dark energy facilities, particularly at stage IV, where systematic
error management is built deeply into the design, will provide many
precise tests and will characterize the behavior of dark energy
beyond merely a single parameter value and its evolution
with time.  We will know the strength of the effects in a
two-dimensional parameter space of distance and cosmic time, as well
as any deviations from expectations in their correlations.
Further surprises may await us.

Inflation is the leading paradigm for the dynamics of the very early
Universe, and current observations of large-scale structure lend
support to this intriguing idea.  The most direct available probes of
inflation come from CMB observations, and the overall agreement is
remarkably good.  However, it has not been possible to explore the
underlying physics of inflation until now:  The coming generations of
CMB experiments will have sufficient sensitivity to falsify large
classes of models.  The signal is a characteristic pattern with
non-zero curl (called ``$B$ mode''), faintly imprinted on the
polarization of the CMB fluctuations, due to gravitational waves
produced during the epoch of inflation.  The shape of the potential of
the scalar field driving inflation directly affects the spectrum of
gravitational waves and hence the strength of the imprint, $r$ (the
ratio of tensor to scalar power), over characteristic angular scales
on the sky.  The current generation of experiments is sensitive to $r
\sim 0.1$, but over the next 10 to 20 years, improvements of two
orders of magnitude are possible by scaling the number of detectors by
similar factors, from $\sim 10^3$ (current) to $\sim 10^4$ (generation
III) to $\sim 5\times 10^5$ (generation IV).  This would require a
change from the way things have been done in the past.  Groups would
merge into one coordinated effort, tapping national lab facility design,
integration, computing, and management capabilities.

In addition, 
 future optical and CMB cosmic surveys, as well as future
polar-ice neutrino projects (see below), will  provide precise
information about neutrino properties, including the mass hierarchy,
the number of light neutrinos, and the sum of the neutrino 
masses.  Combining this with  information from  accelerator- and
reactor-based neutrino experiments, as well from 
experiments searching for neutrinoless double-beta decay, will accelerate
our understanding of fundamental neutrino properties and enable us to
understand the implications of apparent inconsistencies.

Snowmass provided an excellent opportunity to address common problems
and to develop a common vision for the potential of cosmic surveys to 
advance
particle physics.  Highlights included developing detailed strategies
to distinguish dark energy from modified gravity; exploiting the
complementarity of probes for determining the key cosmological
parameters; understanding more deeply the strengths and ultimate
limitations of the different techniques; and discussing the planned
facilities, which are the result of intensive community processes over
many years.  The group articulated a set of goals: (1) remain a leader
in dark energy research, (2) build a generation IV CMB polarization
experiment, and (3) extend the reach of cosmic surveys with targeted
calibration campaigns, targeted R\&D, and support for work at the
interface of theory, simulation, and data analysis.

{\bf Cosmic particles:}
Measurements of fluxes of cosmic particles (charged particles,
photons, and neutrinos) also address many topics in particle physics
beyond indirect dark matter searches.  Recent results include the
detection by IceCube of very high-energy neutrinos that are likely to be from
astrophysical sources; the observation of the GZK suppression in the
cosmic-ray flux above $3\times 10^{19}$ eV; the measurement of the positron
fraction up to 300 GeV, suggesting the existence of primary sources of
positrons from astrophysical processes and/or dark matter
interactions; and confirmation that supernova remnant systems are a
source of galactic cosmic rays.  These and other discoveries were made
by the current generation of experiments.

Goals for the coming decade include determining the origin of the
highest energy particles in the Universe, measuring interaction
cross sections at energies unattainable in terrestrial accelerators,
detecting the GZK neutrinos that arise from the interactions of
ultra-high-energy cosmic rays with the CMB, determining the neutrino
mass hierarchy, and searching for other physics beyond the Standard
Model.

To meet these goals, the group recommends: significant
U.S. participation in the Cherenkov Telescope Array (CTA), which is
the next-generation ground-based gamma-ray facility; simultaneous
operation of Fermi, HAWC, and VERITAS, the current generation of
space- and ground-based U.S.-led gamma-ray facilities; construction of
the PINGU neutrino detector to lower the energy threshold to a few GeV
and enable the determination of the neutrino mass hierarchy using
atmospheric neutrinos; continued operation of the Auger and Telescope
Array air shower arrays with upgrades to enhance the determination of
the composition of the flux of cosmic rays around the GZK suppression
region; construction and deployment of the JEM-EUSO mission aboard the
International Space Station to extend observations of the cosmic ray
flux and anisotropy well beyond the GZK region; and construction of a
next-generation ultra-high-energy GZK neutrino detector, which will
either detect GZK neutrinos (and constrain the neutrino-nucleon cross
section at ultra-high energy) or exclude all but the most unfavorable
parts of the allowed parameter space.  A detailed census of present
and proposed cosmic particle measurement facilities was produced
during the Snowmass process.

{\bf Conclusions:}
In synergy with the other frontier areas, the Cosmic
Frontier provides to particle physics clear evidence for physics
beyond the Standard Model; profound questions of popular interest;
frequent new results; surprises, with broad impact; a large discovery
space with unique probes; important cross-frontier topics; and a full
range of project scales, providing flexible programmatic options.  For
each area of the Cosmic Frontier, the Snowmass study identified
essential technologies and facilities, the advances required in
theoretical models, and experiments with great promise.  The largest
projects are, appropriately and necessarily, international.  The
U.S. is still the leader in many areas of the Cosmic Frontier, but
this field is evolving quickly and other regions with intensive
interest in this physics are advancing rapidly.

\section{Theory}
\label{sec:theory}

This section summarizes the report of the Snowmass Theory Panel.  The DPF
constituted this panel with the goal of understanding both the scientific
problems and opportunities of the next decade, and the challenges involved
in sustaining a first-class theory program in the U.S.

Theoretical physics has played a crucial role in particle physics since its
earliest days.  Theorists developed the basic framework in which we
understand elementary particles:  quantum field theory.  This framework
embodies Einstein's principles of special relativity and locality of
interactions within the laws of quantum mechanics.  It is extraordinarily
successful.  Theorists appreciated the role of symmetries as organizing
principles for understanding data and clues to the nature of physical law.
They developed calculational methods for quantum field theories, permitting
the computation of scattering amplitudes, bound state masses, and numerous
other quantities, often with extremely high precision.  These developments
combined to  both produce and test the Standard Model.  

The discovery of a scalar particle at the LHC may well mark the completion of
the Standard Model.  This object is likely the Higgs boson of the simplest
version of the theory.  Theorists have played and will continue to play
essential roles in firmly establishing the identity of this object.  Its study
at colliders requires not only great experimental ingenuity and persistence,
but also an array of theoretical tools for calculating the rates for its
production and decay.  Just as crucial are techniques for the calculation of
the large backgrounds arising from other Standard Model processes.

While theoretical studies of quantum field theory and the SM
have a long history, two areas have seen extensive progress in the last
decade, and will continue to be the subjects of intense development.  The
first of these is perturbative methods for the calculation of scattering
amplitudes.  In recent years, calculations essential for collider physics,
and previously believed essentially impossible, have been carried through
using a range of new methods.  These computations played a crucial role in
the discovery of the Higgs boson, and are vitally important
in searches for new physics.  The development of new methods, and their
application in LHC and other experiments, as well as in theoretical
investigations, will remain a major activity in the field in the coming
years.

Another area of striking progress has been lattice gauge theory.  This is
the principal tool we have for the quantitative study of the strong
interactions in processes at low energies.  It is now possible to compute
the spectrum of hadrons with high accuracy, and lattice computations have
been crucial in the measurement of the properties of heavy quarks.
Continuing improvements in calculational methods are anticipated in coming
years.  At Snowmass we heard about new applications of lattice methods, such
as prospects for computations essential to any convincing interpretation of
results from the upcoming muon $g- 2$ experiment.

When we say the Standard Model may now be ``complete''
we mean that the theory is consistent to much higher energies.
But there are strong reasons to believe that, at energies not much higher
than those we probe today, there should be new phenomena.  Theorists have
been the drivers in formulating these questions, and proposing possible
solutions.  Among the questions are:
\begin{enumerate}
\item  Why are there vastly
disparate mass scales in nature, such as the Planck mass and the weak scale?
\item  Why are neutrinos light?
\item  Are neutrinos their own antiparticles?
\item  What is the origin of the asymmetry between matter and antimatter?
\item  What is the identity of dark matter?
\item  What is the identity of dark energy?
\item  What is the origin of the curious pattern of quarks and leptons,
and their masses?
\item  Do the forces unify?
\item  What modifications of our basic understanding are required to
reconcile quantum mechanics and gravity?
\end{enumerate}

For each of these questions, theorists have proposed answers.  Many of these
are the subject of present or planned experimental searches.

The Planck scale and the weak scale 
differ by at least 15 orders of magnitude.  The difficulty of
explaining 
the existence of such widely different scales is called the
``hierarchy
problem.''   The large ratio of scales might be 
viewed as simply a ``fact,'' but within quantum field theory, this sort of
hierarchy is generally quite unstable.  (This is usually referred to as the
``fine tuning'' or ``technical naturalness'' problem.)   Proposals to solve
this problem all suggest physics at or near the TeV scale.  Among the most
explored of these is supersymmetry, a possible new symmetry of nature,
which connects fermions and bosons.  The LHC is actively searching for the
new particles predicted by the supersymmetry hypothesis, and has excluded
many popular models.  Alternative proposals include the possibility that
the Higgs particle is composite, or associated with phenomena in dimensions
of space-time beyond the usual four.  Over the next decade theorists will
continue to explore these and other models, incorporating the constraints
from experiment, or the results of discoveries.

We know that the Universe consists largely of forms of energy not contained
within the SM:  dark matter and dark energy.  The case for
dark matter, a form of matter which behaves, essentially, like dust and
interacts extremely weakly with ordinary matter, has become compelling
in the past decade.  Theorists have proposed several persuasive ideas
for what the dark matter might be and how it was produced in the big bang.
Supersymmetric models, in fact, naturally yield candidates (so-called weakly
interacting massive particles, or WIMPs), which are automatically produced
in roughly the right quantities.  These particles are the subject of active
experimental search at accelerators, deep underground, and in space.
Theorists are working actively to survey the possible models and to
understand exclusions and possible signals in ongoing and future experiments.  

But there are other candidates for the dark matter.  Perhaps the most
prominent of these is the axion.  This particle was proposed to explain
perhaps the largest remaining puzzle of the strong interactions:  the
conservation of CP.  The equations of QCD include 
 a parameter, $\theta$, a pure number,
which violates CP.  Exquisite experiments set a limit on the size of any
electric dipole moment of the neutron, and this, in turn, requires $\theta
< 10^{-10}$.  One possible explanation for this small number is a new particle,
called the ``axion", whose dynamics adjust $\theta$ to a value close
to zero.
It turns out that if the axion exists, it is also a candidate for the dark
matter.  Theorists continue to refine the axion theory, exploiting developments
in field theory and in string theory, and to explore its properties.  The ADMX
experiment at the University of Washington is currently searching for this
particle and has a good chance to find it if it exists.

Dark energy is equally mysterious.  Representing about 70\% of the energy
budget of the Universe, this substance has {\it negative} pressure.  Most
theorists suspect that this is Einstein's ``cosmological constant", and the
data to date are consistent with this interpretation.  But its value is very
puzzling.  Conventional ideas of quantum field theory suggest that there
should be much more of it, and even more puzzling is the fact that its
density is just such that it is becoming important in the current epoch of
the Universe.  These questions occupy the attention of many theorists, and
there are a number of proposed answers, but it is safe to say that there is no
compelling picture, and that this will certainly remain an active area of
theoretical investigation for some time.

Neutrinos are now known to have mass, and we know some features of their
masses (mass matrix).  Neutrinos are far lighter than other particles, and
their masses are not accounted for within the SM itself.
Theorists have identified two possible mechanisms to generate neutrino mass.
One is associated with new particle.  These
particles might have enormous masses.  In this case, neutrinos are their own
antiparticles.  Alternatively, there might be extremely light additional
degrees of freedom.  This is a question that can be tested experimentally,
and which has several theoretical consequences.  Understanding the neutrino
masses and mixings will be a central part of both the experimental and
theoretical particle physics programs over the next decade.

Within the  5\% of the energy budget which consists of ordinary protons and
neutrons (baryons), there is a further puzzle:  why is there matter at all,
i.e., why didn't the Universe emerge from the Big Bang with equal amounts of
matter and antimatter?  With the discovery of CP violation 50 years ago,
it was recognized (first by Andrei Sakharov) that this is a question
that can be addressed by science.
  Theorists have understood that the SM, however,
does not violate CP sufficiently to account for the observed asymmetry;
additional degrees of freedom (particles) are an essential component.
They have put forward a number of proposals for how the asymmetry might
arise.  Some of the most compelling lie within the
frameworks of theories of lepton mass (``leptogenesis") and supersymmetry.
These ideas might have observable consequences for the cosmos, such as the
emission of gravitational waves, and for experiments at accelerators.

Much of the history of particle physics has been tied up with the problem of
``flavor":  the existence of three generations of quarks and leptons, and
the features of their masses and mixings.  Theorists have been central to
this subject, proposing the idea of quarks and explaining the problem of
mixing.  In recent decades, they have developed theoretical tools to
understand the behavior of heavy quarks, and a range of ideas for how the
repetitive structure of quarks and leptons might emerge.  These include ideas
involving new symmetries, grand unification, string theory and extra
dimensions.  While many of these ideas are plausible, none are, as of yet,
compelling in themselves, and these issues --- dealing with the dynamics
of heavy quarks and seeking an understanding of the basic issues of
flavor --- will 
be the focus of important activities in the next decade.

The unification of forces is a long-standing dream.  In the past few decades,
theorists have put forward concrete ideas about how this might arise and
proposals for experiments that could test the possibility.  With
supersymmetry, quite remarkably, the gauge couplings of the SM
unify at a high energy scale, and many proposals have been put forward for
an underlying explanation.  A simple and compelling set of ideas of this type
go by the name ``grand unified theories."  These elegantly enlarge the structure
of the SM.  The most dramatic consequence of all of these
proposals is the prospect of proton decay, which has been the subject of
extensive experimental search.
Ideas of unification have also led to a rich set of theoretical questions,
including the existence of magnetic monopoles.

Beyond grand unified theories, the most ambitious attempt to unify the forces
is associated with ``superstring theory."  What is called string theory is
part of a larger, only partly understood, structure which unifies Einstein's
general relativity and the other known forces in a quantum mechanical
framework.  While many questions are not yet answered, string theory has
provided insight into longstanding questions in particle physics, including
the unification of forces, the strong CP problem, dark matter, and dark
energy. It has also inspired much interesting phenomenology, such as that associated
with large extra dimensions.  This is an area that will continue to occupy
a significant fraction of the community in the coming years, and which is
likely to see significant additional progress.

Much of the panel's effort was devoted, as required by its charge, to
examining structural issues in theoretical physics.  The panel's report deals
at some length with questions of funding.  The panel was concerned that the
current budgetary climate at both NSF and DOE puts in jeopardy the research
program that we have outlined above.  Most prominent among our concerns was
support for postdoctoral fellows and students.  Students and postdocs are
important drivers of research, and clearly represent the future of the field.
The panel recommends keeping the current level of support for productive
research groups, of roughly one postdoc and one graduate student per two PIs.

The panel supports the program of comparative reviews, recently introduced by
the DOE.  This permits the agency to look critically at the support levels of
individual theory groups, moving away from a model where funding levels were
usually determined by making modest adjustments to historical levels of
support.  It permits the funding of new research groups and dropping 
groups which have become less productive (as the
NSF has done historically).  This is
essential to adapting to the present funding climate.

The panel understands the need to increase the fraction of the DOE budget
devoted to projects, but argues that this has particularly severe consequences
for theoretical physics.  We proposed that the DOE consider a project category
aimed at theory, and in particular designed to sustain a suitable population
of postdoctoral fellows.  One suggestion is the creation of ``theory
networks," loosely modeled on networks established in Europe.  The DOE would
call for proposals to compete to establish such networks, with a lifetime of
three to five years.  The central topics of investigation would be determined
by the institutions, but we envision  that they might range from intense,
phenomenological efforts in areas like neutrino physics, to investigations
of more foundational issues in field theory and string theory.

\section{Accelerator Capabilities}
\label{sec:AccCap}

The Accelerator Capability study is a synthesis of individual workshops of six
working groups plus the collective Snowmass meeting of all interested
participants. Each group addressed major challenges foreseen for their
respective class of accelerators in a pre-Snowmass meeting. The groups also
considered a set of “big questions” regarding accelerator capabilities for the
long-term future of high energy physics:
\begin{enumerate}
\item How can one build a collider at the 10 -- 30 TeV constituent mass scale?
\item What is the furthest practical energy reach of accelerator-based
particle physics?
\item How would one generate ten or more megawatts of proton beam power? 
\item Can multi-megawatt targets survive? If so, for how long?
\item Can plasma-based accelerators achieve energies and luminosities relevant
to particle physics?
\item Can accelerators be made an order of magnitude cheaper per GeV and/or
per MW?
\end{enumerate}
The results of the workshops formed the basis for draft reports from the
working groups that were discussed in the general Snowmass meeting to form
this consensus summary.  

{\bf Hadron colliders: } 
This working group focused on the evolution of the LHC and possible designs for
a (much) higher energy proton collider (VLHC). The group considered: (1) how high
a luminosity is possible for the LHC, (2) what are available increasing
integrated luminosity without compromising experiments or detector survival,
(3) how high an energy is possible in the LHC tunnel,  (4) what 
impediments exist to designing a 100 TeV collider, and  (5) what the associated accelerator
research roadmap should be for hadron colliders.

The priority recommendation of our study is full exploitation of the  LHC.  Doing so
requires a strong LHC Accelerator Research Program sponsored by the
Office of
High Energy Physics that
transitions to a US-LHC high luminosity construction project. During the
project period we recommend continuing a focused, integrated, laboratory
program that emphasizes the engineering readiness of technologies suitable for
a 26 TeV upgrade of the LHC or a machine of higher energy in a larger tunnel.
The most critical technology development toward higher-energy hadron colliders
is the next-generation high field Nb$_3$Sn magnets (limited to ~15 Tesla) and
adequate beam control technology to assure machine protection.

The reach of an LHC energy upgrade is constrained by the limits of Nb$_3$Sn
technology and by the absence of engineering materials with high-field
properties beyond those of Nb$_3$Sn. Moreover, even doubling the LHC energy in the
present tunnel introduces substantial issues of synchrotron radiation
management. Radiation management will become very difficult as the synchrotron
power on the beam tube reaches 5 W/m.

To achieve energies beyond those of the LHC, the multi-laboratory study of
VLHC
 remains valid.  Snowmass
has stimulated renewed effort on the VLHC in  both the U.S. and Europe.  American
participation in the CERN-led international study for colliders in a large
tunnel that will begin in 2014 will inform decisions to expand the
reach of U.S. technology
and guide research investments. The areas in which U.S. accelerator scientists
can make the most valuable contributions are beam dynamics, superconducting
magnets, vacuum systems, and machine protection.

Long-term, innovative research will expand the technical options for any future
hadron collider. Dipoles with operating fields beyond 15 T need new 
conductor elements such as small-filament, high-temperature superconductors in
continuous kilometer lengths.  Better conductors, innovative stress management,
and novel structural materials will enable even higher-field magnets with
greater temperature margin.  With ever more stored energy in the beams, better
understanding and modeling of beam dynamics is essential to control beam halos
and lost beam particles. Machine protection and design of beam abort dumps for
multi-GJ beams will be challenging. Other issues for research include effects
of marginal synchrotron radiation damping, beam physics of the injection chain,
effects of noise and ground motion, and options for interaction region design.

{\bf Energy-frontier lepton and photon colliders:}
Our study welcomes the initiative for the International Linear
Collider (ILC) in Japan. The ILC would begin as a 250
GeV Higgs factory with future expansion to 500 GeV. The U.S. accelerator
community is capable of contributing to the ILC
 as part of a balanced U.S. particle
physics program. As described in its Technical Design Report (TDR), the ILC is
technically ready to proceed to construction. The TDR incorporates leadership
U.S. contributions to machine physics and technology in superconducting RF
(SRF), high-power targets for positron production, beam delivery, damping ring
design, and beam dynamics such as electron cloud effects. Extending the ILC to
1 TeV would require lengthened linac tunnels and added cryomodules. It would
use the original ILC sources, damping rings, beam delivery systems, and beam
dumps.

The excitement surrounding the Higgs boson discovery stimulated consideration
of alternatives to a SRF Higgs factory. Concepts include a linear collider
using copper linacs, a large-circumference $\ee$ ring, a compact muon collider
ring, a photon collider, and $\ee$ linear colliders based on wakefield
acceleration techniques.  These concepts span a broad range of technical
readiness (from requiring demonstration of feasibility to having a detailed
conceptual design) and timescales upon which a machine could be constructed.
They also have varying energy reach from the hundreds of  GeV scale to the
multi-TeV regime.   

It is natural to investigate  whether a 250 GeV Higgs factory could fit in the LHC
tunnel. This option is undesirable because it interferes with LHC operations
and because the beam physics is highly constrained. Assessment of a circular
collider in a very large (of the order  of 100 km)  tunnel, with an energy reach up to about
400 GeV,  will be part of the CERN-led study of large colliders
mentioned above. Such a machine is a
substantial extrapolation from existing and past storage rings, albeit from a
large experience base. Beamstrahlung at the interaction point strongly couples
energy reach and luminosity. The luminosity would be largest at the $Z$ peak,
but fall rapidly as the center of mass energy increases. Should the ILC not go
forward over the next decade and should the renewed interest in a very large
circumference hadron collider be sustained, the possibility of a circular Higgs
factory deserves extensive consideration. 

In a Higgs factory photon collider, two electron beams are accelerated to 80
GeV and converted to ~63 GeV photon beams via inverse Compton scattering
against low-energy (3.5 eV), high-intensity (5 J) laser pulses. The high-energy
photon beams then collide to generate Higgs bosons through the $s$-channel
resonance, $\gamma\gamma\to H$.  A photon collider has the distinct advantage
of requiring only an 80 GeV electron beam energy. Photon colliders could
accompany proposed linear or circular colliders or be  stand-alone facilities.
The laser technologies overlap with those for laser wakefield accelerators.

Muon accelerators could provide world-leading experimental capabilities at
energies from the Higgs $s$-channel threshold at 126 GeV up to the multi-TeV
scale. A circular muon collider, if feasible, could reach such energies,
because the larger mass of the muon suppresses synchrotron radiation. As muons
at rest have a lifetime of 2.2~$\mu$s, they will decay in flight. The short muon
lifetime demands that beam creation, manipulation and acceleration to high
energy be done rapidly; high-gradient acceleration is essential. An 
Energy
Frontier muon collider would necessarily be relatively compact. Even a 5 TeV
collider would fit on the Fermilab site. Critical beam physics issues are: (1)
cooling the muon phase volume by $10^6$, and (2) accumulating $10^{12}$ $\mu^+$
and $\mu^-$ bunches in the collider.  A vigorous, integrated R\&D program
toward demonstrating feasibility of a muon collider (Muon Accelerator Program)
is highly desirable. The current funding level is, however, insufficient for
timely progress. Development of a muon collider capability would be closely
connected with Intensity Frontier accelerators such as intense neutrino
sources.

The Compact Linear Collider (CLIC) concept is based on 100 MeV/m copper linac
technology; it would stretch 50 km for a 3 TeV collider. CLIC would be powered
by two high-current drive beams running parallel to the colliding beams through
a sequence of power extraction and transfer structures, where they produce
short, high-power RF pulses that are transferred into the accelerating
structures. The practical energy reach depends on control of wakefields and on
the accelerating gradient in industrialized accelerator sections. U.S. national
laboratories have substantial expertise in CLIC technologies.

Yet another approach for multi-TeV energies proposes to use wakefields in
plasmas or dielectric structures driven either by beams or lasers to achieve
accelerating fields of 10 to 100 GeV/m. Many feasibility and practicality
issues remain: positron acceleration, multi-stage acceleration, control of beam
quality, and plasma instabilities at tens of kHz repetition rate. All variants
require an integrated proof-of-principle test. The U.S. is a world leader in
these strong physics programs at the frontier of accelerator science.

{\bf High-intensity proton sources for neutrinos, muons, and rare processes:} 
Requirements for Intensity Frontier experiments are more diverse than for
the Energy Frontier.  Therefore, this study addressed a set of structured
questions: (1) What secondary beams are needed for Intensity Frontier experiments? (2) What
proton beams could generate such beams? (3) Can these proton beams be made by
existing machines? (4) What new capabilities are needed? (5) What accelerator
and target research is needed to realize the new capabilities? 
The study
surveyed particle physics requirements for secondary beams, including
beams of  neutrinos,
kaons, muons, and  neutrons.   Experiment advocates supplied nineteen secondary
beam requests. From these the study group derived primary proton beam
characteristics. 

The common characteristics required are average beam power, with more than 1 MW
delivered, and a flexible, experiment-dependent time structure. Beam
requirements were compared with 20 existing proton beam-lines and 14
planned upgrades. The overarching conclusion is that the next generation of
intensity frontier experiments requires beam intensities and timing structures
beyond the capabilities of any existing accelerator.

Fermilab's proposed multi-stage Project X would yield a world-leading facility
based on a modern multi-MW superconducting proton linac capable of injecting
into the Fermilab Main Injector.  The linac would deliver a flexible
“on-demand” beam structure that could serve multiple experiments over an energy
range 0.25--120 GeV. The linac would provide a platform for future muon
facilities including nuSTORM, a neutrino factory, and a muon collider. A
complete, integrated Reference Design Report identifies technical risks that
will be mitigated in a structured research program already underway.

 The DAE$\delta$ALUS collaboration proposes
multiple sources of decay-at-rest anti-neutrinos for short-baseline oscillation
experiments. This project has narrower experimental scope than Project
X.  DAE$\delta$ALUS would use three multi-MW H$_2^+$ cyclotrons and target
stations located about 2 to 20  km from a large hydrogenous detector.
The experiment would measure CP
violation in a way that is complementary to the LBNE experiment. The first stage of
DAE$\delta$ALUS would be  IsoDAR, a compact 60~MeV cyclotron located only 15~m from the
KamLAND detector, that  would make a definitive search for one or two sterile neutrinos.
This international collaboration has strong connections with commercial
cyclotron industries.

Another possibility is nuSTORM, neutrinos from STORed muons.
This would be a first
step toward a long-baseline neutrino factory capability. The nuSTORM muon
storage ring would send well-characterized neutrino beams to detectors at 50 m
and 1900 m for a sterile neutrino search and neutrino cross-section
measurements. 

A common research issue for Intensity Frontier  capabilities is the injection system,
composed of low-emittance, high-current ion sources with effective beam
choppers. Control of space-charge forces is important for preserving beam
quality.  Understanding and limiting beam loss is a dominant operational issue
requiring adequate simulation of halo formation, efficient beam collimation,
and very high-efficiency extraction.
 
High-power targets are a difficult challenge that limits facility performance.
The principal underlying damage mechanisms of the target materials are atom
displacements and gas production. Particulars depend on primary beam
characteristics, target material, operating temperature, and the duty factor of
the accelerator. Unfortunately, one cannot directly translate
experience
with nuclear reactors to estimate the performance of targets with 
high-energy beams from experience with nuclear reactors. Details of target
behavior and failure mechanisms are a mesoscale problem that is difficult to
simulate. Computed radiation effects in inhomogeneous materials subject to
time-varying irradiation need validation with controlled, instrumented in-beam
tests. 

{\bf High-intensity electron and photon beams:} 
This working group addressed two major questions: (1) What capabilities at heavy
flavor factories are required to realize the full range of physics
opportunities? (2) What are new physics opportunities using high power electron
and positron beams?  The relevant technologies exploit strong synergy with
light sources and damping rings.

SuperKEKB, a super-high-luminosity $B$-factory, is an upgrade to the KEKB
B-factory currently under construction in Japan with commissioning to commence
in January 2015. To achieve the target luminosity of $8 \times 10^{35}$
cm$^{-2}$s$^{-1}$, a forty-fold increase over that of KEKB, the SuperKEKB
beam currents will be approximately twice as high as used at KEKB, and vertical
bunch sizes at the collision point about 20 times smaller than those achieved
at KEKB. Greater U.S. collaboration would strengthen the SuperKEKB project and
might enable even higher luminosity well in excess of $10^{36}$~cm$^{-2}$s$^{-1}$.

Two super tau-charm factories have been proposed: one at Frascati (Tor Vergata)
in Italy and one at Novosibirsk in Russia. Both machines are two-ring,
symmetric-energy machines, with provisions for longitudinally polarized beams. 

The DarkLight experiment will use the high-intensity electron beam of the JLab
FEL, impinging on a hydrogen target, to search for gauge bosons associated with 
``dark force'' theories. It might also be possible to use an intense, low-emittance
positron beam, impinging on a plasma target, to generate sufficient
muon/anti-muon pairs to provide a source beam for a future muon collider
without the need for a separate muon cooling stage.

{\bf Electron-ion colliders:} 
Several future electron-ion colliders have been studied in recent years. All
would be based at an existing accelerator facility. The collider configurations
include both ring-ring and linac-ring options. Center of mass energies range
from 14 GeV to 2000 GeV.  Most of the collider concepts share several enabling
technologies. SRF cavities must be able to operate with high
average and high peak beam currents, providing effective damping of high-order
modes. The cryomodule design must be consistent with containing high beam
power. Hadron beam transverse and longitudinal emittances must be small
to achieve high collider luminosity. Therefore, the designs with medium hadron
energy call for the application of powerful cooling techniques.

The low $\beta^*$ interaction region designs for all proposed colliders
require strong focusing of beams at the collision point and fast
separation of beams after the collision. The synchrotron radiation fan produced
by electrons in the focusing magnets must be kept from hitting the
the beam pipe in 
the vicinity of the detectors and inside  superconducting magnets. 
 
The linac-ring designs utilize a polarized electron source, with an average
current ranging from 6 mA to 50 mA. The linac-ring scheme introduces
non-standard beam-beam effects, which must be explored to understand the limits
on the luminosity and the beam parameters.  Other shared technologies include
techniques to preserve beam polarization. Spin matching
and the harmonic correction techniques have to be investigated for 
ring-ring colliders  to minimize the
beam depolarization due to synchrotron radiation, especially in the presence of
spin rotators and solenoidal detector magnets.

{\bf Accelerator technology test beds:} 
We identified a broad range of existing and needed test capabilities for
proposed frontier accelerators. The first category of test facilities permits
testing beam physics or accelerator components essential to manage technical
risks in planned projects. A second category would integrate accelerator
systems to provide proof-of-practicality tests.  The third category provides
tests of physics feasibility of concepts and/or components.  This study
identified 35 existing facilities in the U.S. and overseas, both with and
without beam testing capability. Although these facilities provide substantial
readiness to move forward with the highest priority accelerators for
particle physics, the long range future of particle physics needs a few
additional dedicated test capabilities in the near term.

\section{Underground Laboratory Capabilities}
\label{sec:underground}

Some of the most compelling experiments in particle physics can only be done at underground facilities. Searches for dark matter and neutrinoless double beta decay and neutrino experiments using solar, reactor, atmospheric, and supernova neutrinos and neutrinos from accelerators all require underground facilities and capabilities. 

Underground facilities are located in North America, Europe, Asia, and Antarctica (in ice). New underground facilities have become operational in all of these regions in the last few years.  The world-wide particle physics community plans to expand underground capabilities over the next years, primarily 
outside the United States. If all of these plans are realized, general-purpose space for underground experiments will roughly double by the end of the decade.  The expansion would include major new facilities 
to host reactor experiments at moderate depths, and a new class of very large facilities for long-baseline 
and atmospheric neutrino experiments, proton decay, and other physics.
 
Plans for expansion or continuation of underground facilities in the United 
States are less developed. Currently, there are no plans with approved federal
 funding for expansion of 
underground capabilities at the four underground sites located in the United 
States. The Long-Baseline 
Neutrino Experiment (LBNE) has provisional approval to be located on the 
surface at the Sanford Underground Research Facility (SURF) 
in South Dakota, but design work is underway in anticipation 
of achieving a global collaboration to allow LBNE to be sited deep 
underground at SURF. The LBNE physics community expressed strong
 support for the deeper site during the Snowmass process.

All of the next generation (G2) dark matter experiments can be accommodated by
 existing or planned underground facilities, assuming no reduction in these facilities 
for the rest of the decade. Only one of these experiments is planned to be
 located at a U.S. facility.
 Several neutrinoless double-beta-decay experiments are already under construction at 
existing underground facilities, one of which is in the U.S.   Next-generation (ton scale) 
neutrinoless double-beta-decay experiments can likely be accommodated by existing and
 planned facilities, but will face competition for underground laboratory space from dark
 matter experiments. 

Detectors for reactor experiments with baselines greater than 100 m require 
medium-depth underground laboratories. Future reactor experiments are being 
planned overseas based on funding 
commitments from the host countries. 

The flagship of the international neutrino effort is the search for CP violation
 in the lepton sector, which requires a massive detector and a very intense neutrino beam. 
There are other motivations for 
constructing this 
massive detector underground.
The search for nucleon decay is one of the most important topics in 
particle physics. Atmospheric neutrinos, observable in a large underground detector, 
may be sensitive to all of the currently poorly known  neutrino oscillation parameters.
 The spectacular neutrino burst from a nearby supernova event would be detected 
at no additional cost if the detector is underground, but such detection is very 
difficult on the surface. Some of the same detectors that would be used for 
long-baseline neutrino experiments could be used to advance the 
search for CP violation, nucleon decay, the study of atmospheric neutrinos, and other
 physics if the detector is located underground. This is the plan for Hyper-K (Japan) 
and LBNO (Europe). It would be a lost opportunity if this condition cannot be satisfied
 with LBNE.

Experimental needs for materials assay and storage outstrip the capability of existing facilities, and space for such work should be reserved at new facilities. In addition, underground space should be reserved for small prototype testing and generic R\&D. There is enough space at U.S. facilities to meet  future needs if the existing underground labs are maintained.

As the scale and cost of underground experiments grows it will become even more important to maintain open competitive access to underground laboratories. The best way for the governments to support the international system of underground experiments is for each major country (or region) to support at least one major underground laboratory capable of hosting forefront experiments. It is not clear whether it would be possible to sustain this international support if one country chose to take a major role in the research without supporting any facility. 

Our conclusions are:
\begin{enumerate}

\item{We should locate LBNE underground to realize its full science potential. This step would also provide a natural base for additional domestic underground capabilities at SURF in the future.}
\item{The U.S. has leading roles in many of the future dark matter, neutrinoless double-beta-decay and neutrino experiments.} 
\item{More coordination and planning of underground facilities (overseas and domestic) is required to maintain this leading role, including use of U.S. infrastructure.}
\item{Maintaining an underground facility that can be expanded to house the largest dark matter and neutrinoless double-beta-decay experiments would guarantee the ability of the U.S. to continue its strong role in the worldwide program of underground physics.}

\end{enumerate}
\section{Instrumentation }
\label{sec:instrum}
 
The search for answers to fundamental questions in the field of particle physics has always been 
intimately tied to the development of  innovative technologies or significant 
advancements in existing technologies.  
The particle physics community has a long history of inventing detectors based on new technologies to 
address the science needs and advancing these technologies to large-scale reliable use. 
For many decades the 
instrumentation needs of particle physics have motivated university faculty, 
national laboratory staff, and industrial scientists to develop the technological 
foundations and invent the detectors responsible for many of the important 
particle physics discoveries. The field of particle physics is generally regarded 
as an incubator of innovation in instrumentation.  Moreover, driven by the needs of particle
physics, the technology developed for accelerators and detectors has historically 
benefited many other fields of the physical and applied sciences and medicine. 

Modern particle physics experiments and high-energy accelerators put extraordinary 
demands on sensors, sensor readout electronics, precision engineering, and data acquisition 
and management, often incorporated into detector systems of very large scale. 
In addition, the scientific approaches are broadening to include lower-energy, high-intensity 
and ultra-low-background experiments, including experiments deploying very large volume detectors 
to study very rare processes, and also experiments that study fundamental properties of the 
cosmic energy and matter with greater precision. 
 
Focusing only on scaling up existing technologies to larger experiments or carrying out 
detector R\&D only when it is needed is tempting in tight budgetary times, but this 
is counter to the successful approach that has been followed up to now. 
Instead, more fundamental innovation and development of new approaches are necessary 
to make experiments feasible or economically viable. 
An appropriate investment in a detector R\&D program will be required to enable the science 
goals and meet the budgetary challenges.  This program would 
develop,  over intermediate
and
long time frames,  new tools and technologies that are both 
cost-effective and have an enhanced physics reach.
It will allow the field to continue to carry out flagship domestic experimental 
research and have leadership roles in off-shore experimental projects, while the 
development of new, transformative detection capabilities will ensure an affordable and 
healthy experimental particle physics research program in the future. 
The major challenge for instrumentation is to structure the current advanced detector R\&D 
program such that it will enable the United
States to continue to maintain scientific leadership in many key areas of a broad international
experimental program in particle physics. 

The instrumentation needs of planned and proposed future experiments across the field were 
surveyed in two joint preparatory meetings, in several dedicated topical workshops and in
joint sessions with the Energy, Cosmic, and Intensity Frontiers during the Snowmass
study. In addition, nearly one hundred white papers on instrumentation were submitted, 
covering a broad range of topics. In all cases instrumentation 
development was considered central to progress. 
An overview of the current and planned programs of experiments at the various physics frontiers
and specific detector needs is provided in the full Instrumentation report. 
This survey of the whole experimental program of particle 
physics  identifies key issues in instrumentation within the next
decade and beyond. It also gives 
a picture of the opportunities that exist to establish an instrumentation program that 
can satisfy both the needs of particle physics and at the same time create an environment 
of innovation to benefit other fields of science.
This summary articulates a vision for instrumentation that will 
enable execution of a broad, targeted experimental program within the fiscal realities of 
our time and identify areas where the U.S. can take a leadership position. 

Ideally, the program  of detector R\&D should include a range
of  projects with various levels of risk and time scales for full
development.
Detector R\&D carried out 
within existing experiments is, by necessity, project-driven and provides a low-risk path to 
relatively incremental improvements to existing technologies.  Detector R\&D that is motivated by 
common needs among various experiments is generally longer-term, can be higher-risk, and adds
value to multiple experimental areas at the same time. This type of R\&D can lead to 
incremental or significant improvements in cost reduction, scientific reach, or both.  
At the highest level of impact and risk is long-term detector R\&D leading to 
transformative changes in cost reduction, 
increase in scientific reach or both, across a significant part of the experimental 
program.  This is the kind of high-risk, high-reward detector R\&D that has the potential to 
lead to scientific breakthroughs. Underpinning these R\&D efforts is the urgent need 
for training the next generation of instrumentation experts, without which there can be
no long-term future.

Major technological advances based on a better understanding of the underlying science are 
also occurring in other scientific disciplines such as materials science, photonics and nanotechnology.  
Many of these advances have the potential to lead to transformational new technologies for 
particle physics detectors.  Particle physics should continue to exploit and pursue  
the technology advances in other experimental scientific disciplines, 
which could contribute to opportunities for 
innovation and the development of transformative technologies for particle physics.  

A healthy national instrumentation program must provide a balance between 
evolutionary and revolutionary detector development while training the next generation of experts. 
We have five recommendations for critical elements of this national program:
\begin{enumerate}
\item Support detector R\&D with clearly identified areas of detector development based on 
the strengths of the community.
\item Achieve an appropriate balance between evolutionary and revolutionary 
detector R\&D, i.e., an appropriate ``portfolio of risk,'' and build expertise in 
innovative technologies that can be applied to the design and construction of novel, 
cost-effective particle physics detectors.
\item Develop a process for optimizing the use of existing university, national laboratory, 
and industrial resources, to grown and retain local technical expertise at 
universities and laboratories, and to identify incentives and mechanisms for 
improving detector R\&D collaborations and equipment sharing among universities, national 
laboratories and industry.
\item  Create opportunities for attracting, and providing careers for, particle physicists 
with interest in,
and outstanding capability for, innovative detector design and
development.  This community
 of experimental physicists will preserve  the  background and skills needed to design and 
build future generations of particle physics experiments.
\item Provide mechanisms for identifying and transferring appropriate technologies developed in
other scientific disciplines to particle physics and for transferring applicable technologies 
developed in particle physics to other science disciplines, such as nuclear physics, 
basic energy sciences, and related branches of science, medicine, and national security. 
\end{enumerate}

The Snowmass process provided broad input and guidance from the
particle physics 
 community in identifying 
these crucial elements of a national technology and 
instrumentation program. A previous DPF Task Force on Instrumentation led to the creation of 
CPAD, the Coordinating Panel for Advanced Detectors,
which is intended to act as the advocate for the detector development
 program, to promote its merits, and to provide venues for 
regular presentation of results.   CPAD can bring different groups of
technical experts 
together to make the community aware of developments elsewhere. 
CPAD can also coordinate between the funding agencies and the instrumentation
 developers by providing information about instrumentation needs and
 ongoing activities.

A crucial enabling element for an instrumentation program is the development of 
expert physicist manpower.  
An investment in younger physicists 
who work on instrumentation is critical to support a long-term
particle physics program.   It is noted that only under exceptional 
circumstances are doctorate degrees awarded by 
U.S. physics department for a Ph.D. thesis based on instrumentation, 
whereas this is commonplace in Europe.
Furthermore, early specialization 
in instrumentation is  strongly disadvantaged given the emphasis on
physics analysis to obtain faculty 
positions. This has to change for the field to remain viable. 
 
Other essential enablers are national laboratory resources, unique facilities such 
as test beams, and targeted funding aimed at a specific problem.
The U.S. national laboratories are a resource of unique importance. 
Their breadth, 
because of their multidisciplinary scientific nature, provides cross-fertilization of 
useful technologies from nuclear physics, basic energy science, materials research,
engineering, chemistry, and computer science. Many U.S. national laboratories have close 
links to top-ranked universities, which are also centers for multi-disciplinary innovation 
and ideas. University groups have greater 
difficulty maintaining long-term technical and engineering resources, 
while the national laboratories naturally maintain these as
consequence of their missions.
By combining the intellectual and
manpower capabilities of universities with the 
resources of the national laboratories 
and the product development capabilities of industry,
the U.S. can continue to confront and 
overcome many of the technological challenges of future particle physics 
experiments. 

The United States has several complementary test beam and 
irradiation facilities. 
These facilities are a critical component of instrumentation work, and there is a separate 
Snowmass report about them. 
The Snowmass process did not allow for a proper evaluation of all facilities for instrumentation, 
and it is recommended that CPAD finalize this process. Adequate support for these facilities is 
essential for a healthy detector R\&D program. 

Another enabler to encourage transformative innovation in instrumentation would be to 
initiate a new program, outside the existing funding, for long-term 
investment in more speculative but potentially high impact research motivated by a set of 
``grand challenges.''
Once identified, such grand challenges would be effective in focusing the creative power
of the intrumentation
community on problems that have the potential for large payoff. Areas where 
existing technologies would be cost-prohibitive for meeting the goals of future experiments 
are good candidates for new initiatives. 
These grand challenges should be issued nationally, and cross-disciplinary collaboration 
should be strongly encouraged. This would allow the program to take advantage of the 
tremendous progress and breakthrough advances that have been made in areas of science 
outside the field of particle physics that could prove very valuable for the development 
of future instrumentation. Because of the cross-disciplinary aspect,  
close collaboration between universities and national laboratories
will be a key component.
Funding would be subject to proposal review, but should be at a substantial level for a 
period of at least three years. CPAD could be engaged to identify the set of grand challenges
that would define the program. 

During the Snowmass Instrumentation discussions, a number of instrumentation 
areas were recognized as of strategic importance. 
These areas all focus on major 
technological barriers that stand in the way of reaching the science goals.  Some of
them have the potential to deliver very cost-effective instrumentation methods and provide  
breakthrough new technology.   The choice of these areas was guided by their physics impact
and existing strengths and capabilities in the country.  Consideration was given 
to the technology's usefulness to other branches of science.  Although some of these 
instrumentation themes seem very challenging, it is likely that many, if not all, can be 
realized with a dedicated instrumentation effort. The main strategic areas that have been identified
are described below.

{\bf ASICs:} The use of Application Specific Integrated Circuit (ASIC) electronics and interconnect 
development is often critical to enable an experiment.
A number of factors make ASICs essential to particle physics, such as small physical size, 
high channel density, ability to integrate a variety of function blocks, low power dissipation 
and radiation tolerance. Examples of  
areas where ASIC-related R\&D is required are high-speed waveform sampling, pico-second timing, 
high-rate radiation tolerant data transmission, low temperature operation, low power and 
2.5D and 3D assemblies. 
The field of particle physics has spearheaded the use of ASICs, but there is a growing need 
and adoption by other disciplines. ASIC development provides an excellent opportunity to work 
more closely with other branches of science by trying to address their instrumentation needs. 
A report summarizing a workshop held earlier in 2013 to look at the
ASIC needs for 
particle physics  was submitted
to the Snowmass proceedings. 

{\bf Calorimetry:} The measurement of the total energy of electrons and jets lies at the core of
experiments at the Energy and Intensity Frontiers. Projects designed to search for very rare 
processes or to make precision measurements are in need of more precise, faster, and more 
cost-effective methods to perform these calorimetric measurements. 

{\bf High-speed data acquisition:} Experiments are required to handle huge interaction rates to acquire, 
transport, process and retain 
the events of interest,  preserve the accuracy of the measurements of intrinsic particle 
properties, and uncover  signatures  of new physics.  More intelligent trigger and data 
acquisition systems are needed to enable higher statistics experiments. 

{\bf Large-volume detectors:} The study of neutrino properties and their interactions 
and the search for dark matter require large-volume detectors at underground facilities. 
Innovative technologies that allow scaling in a cost-effective way with increased sensitivity 
are required to enable the spectroscopy of these fundamental particles. A coherent research 
program in low-radioactive materials and assay is required. 
 
{\bf Photodetectors:} A multitude of physics processes can be studied by measuring photons with 
wavelengths ranging from mm to nm. Instruments used to study these
photons are based on a range of materials ranging from superconductors 
to semiconductors, from alkali metals to crystals. The development of large arrays with 
improved spectral sensitivity, energy and time resolution, and excellent background rejection
would truly revolutionize future experiments. 

{\bf Pixelated sensors:} High granularity has become a requirement for many of our detectors. 
Often the higher density results in performance compromises. The development of new technologies 
designed  to deal with the higher density, while avoiding these compromises and 
improving overall performance, is essential for future experiments. These include sensors with a 
greater degree of pixelation, radiation hardness, high speed, and built-in 
intelligence to carry out a number of operations, including hit time-stamping, clustering  
and recognizing  hit correlations,  that can affordably be deployed in large areas.

{\bf Power and mass:} Especially at the Energy Frontier, experiments are characterized 
by high radiation, huge interaction rates, and serious constraints on power and mass budget.
Better low-mass structural materials that are strong and stable, including materials 
with ultra-low intrinsic radioactivity, would benefit a broad spectrum
of future experiments.  The design of 
electrical power distribution and cooling systems seems mundane, but it
can 
severely limit the physics reach of current experiments. These systems
must deliver services with
low mass in a high radiation and magnetic field environment. Innovative solutions are  
critically important for next generation experiments. 

The particle physics technology and instrumentation program described here requires a multi-year 
commitment from the funding agencies. 
The funding required to meet short-term financial obligations to sustain an existing 
particle physics research program puts enormous pressure on funds 
earmarked for long-term, generic detector development.  In spite of these pressures, a stably 
and adequately funded generic instrumentation program will ensure that the field 
invests in its future and establishes a foundation for a competitive, healthy program in the 
long term.

\section{Computing}
\label{sec:computing}

Computing has become a major component of all particle physics experiments
and in many areas of theoretical particle physics. The Computing 
study group established subgroups covering user needs and infrastructure.
The study considered user needs for  experiments at the Energy and Intensity
Frontiers, and  the combined needs of Cosmic Frontier experiments, astrophysics
and cosmology.  Theory subgroups covered accelerator science,
astrophysics and cosmology, lattice field theory, and perturbative QCD.
Four infrastructure groups examined trends in computing to predict how
technology will evolve and how it will affect future costs and
capabilities. These groups focused on distributed computing and facility
infrastructures, networking, 
data management and storage, and software development, personnel, and
training.
They identified critical
technology needs for particle physics that might require the DOE or NSF to
fund research in computer science and technology.

During the period between the Community Planning Meeting at Fermilab and
the Community Summer Study meeting at the University of Minnesota, the Computing
groups were actively engaged with the other frontiers to learn of
their plans and estimate their computing needs. The infrastructure groups
engaged with vendors, computer users, providers, and technical experts to
predict trends in computing, networking, storage, and software development,
including considerations of costs, capacities, and speeds. Two days of
parallel 
sessions at the Minnesota
meeting were devoted to discussions across
the the subgroups, to finalize subgroup findings, and to identify common
trends and needs.

Progress in particle physics experiment and theory will require significantly more
computing, software development, storage, and networking. Different
projects stretch future capabilities in different ways, but there
are many common needs among the different areas of particle physics.
In the future more commonality and community planning would aid in 
moving 
ahead in the most efficient manner. This requires careful and continuing
review of the topics we studied, in particular,  user needs and capabilities of
current and future technology. For many years, the particle physics
community has been a great source of computing innovation and expertise. It
is essential to leverage those assets through wider sharing of knowledge
throughout the experimental and theoretical communities. 
We should be open to bi-directional sharing of expertise with the entire
scientific community.

The experimental program relies for the most part on distributed
high-throughput computing (HTC). Simulation, data analysis, and
reconstruction of individual events are independent of each other, so that
groups of events can be assigned to hardware in different locations.
Results are combined when all event groups are done. This
distributed computing model was pioneered by the Energy Frontier
experiments.  It relies  on a distributed infrastructure of computing
centers as part of the Open Science Grid in the U.S. and extending across the
globe. Theoretical computing and simulation needs are more commonly
addressed by high-performance computing (HPC), in which thousands to hundreds
of thousands of tightly coupled CPUs are working simultaneously on a single
problem. These resources are provided mostly through DOE and NSF
supercomputing centers.

One issue for those applications that traditionally rely on HTC for their
data-intensive computing is to what degree they can or should use national
supercomputer centers, which have traditionally been designed for HPC usage.
Work is proceeding to make these HTC applications run on HPC, and to
interface HPC centers to the HTC workload and data management
infrastructures. 
Also, traditional HPC applications are developing in the direction of
 more data-intensive
science, which, however, is currently not a good match to existing
and next-generation HPC architectures. Computational resources will have to
address the demands for greatly increasing data rates, and the increased
needs for data-intensive computing tasks like data analytics, for comparing
large samples of simulations and observational data.

Another pressing issue facing both HTC and HPC communities is that
processor speeds are no longer increasing exponentially, as they were for
at least two decades. Instead, new chip architectures provide multiple
cores. Thus, we cannot rely on new hardware to run serial codes faster, and
we must parallelize codes to increase application performance. In addition
to multi-core chips, there are accelerators such as graphical processing
units (GPUs) and many-core chips such as the Intel Xeon Phi. 
Computing resource needs for Energy Frontier experiments used to scale
roughly with the rate that processor speeds increased, following Moore's
law. Future advances will  require full use of multiple-core and many-thread 
architectures. Scaling of disk capacity and throughput is of
significant
concern in the storage area, since per-unit capacities are no longer
increasing
 as rapidly.

These changes in chip technology and high-performance system architectures
require us to develop parallel algorithms and codes, and to train personnel
to develop, support, and maintain them. Different subgroups are at different
stages in their efforts to port to these new technologies. Lattice QCD, for
example, started its GPU porting efforts in 2008 and has had code in
production for some time, particularly for matrix  inversions of the lattice
Dirac
operator; however, there
are other parts of the code that are still only running on CPUs.
Cosmological simulations have exploited GPUs since 2009 and some 
codes have fully incorporated GPUs in their production versions, running 
at full scale on hybrid supercomputers.
Accelerator science is also actively porting codes to GPUs. Some of the
solvers and particle-in-cell infrastructures have been ported and very
significant speed-ups have been obtained. The perturbative QCD community
has also started using GPUs.

These trends lead to vastly increasing code and system complexities. For
example, only a limited number of people in the field can program GPUs. In
this and other highly technical areas, developing and keeping expertise in
new software technologies is a challenge, because  well-trained personnel and
key developers are leaving to take attractive positions in industry.
Continued training is an important aspect. Training materials are now provided
by some of the national supercomputing centers,  and by summer schools
organized by, among others, the Virtual School of Computational Science and
Engineering.  We must examine whether these provide the
right training for our field and whether the delivery mechanisms are
timely. On-line media, workbooks, and wikis are suggested to enhance
training.  Another area of common concern is the career path of those who
become experts in software development and computing. It is useful to help
young scientists learn computing and software skills that are marketable
for non-academic jobs, but it is also important that there be career paths
within particle physics, including tenure-track jobs, for those working at
the forefront of computation.

{\bf Energy Frontier} experiments already experience computing
limitations that limit the amount of physics data that can be analyzed. The
planned upgrades to the LHC energy and luminosity are expected to result in
a ten-fold increase in the number of events and a ten-fold increase in
event complexity. Effort has begun to increase code efficiency and
parallelism in reconstruction software and to explore the potential of
GPUs. The experiments are considering saving more raw events to tape and only
reconstructing  them selectively. The LHC produces about 15 PB of raw data per
year now, but by 2021 the rate may rise to 130 PB.  Attention needs to be
paid to data management and wide-area networking, to assure that network
connectivity does not become a bottleneck for distributed event
analysis. It is important to monitor storage cost and throughputs. More
than half of the computing cost is now for storage, and in the future it
may become cost-effective to recompute certain derived quantities rather
than storing them.

{\bf Intensity Frontier} experiments have combined computing requirements
on the scale of a single Energy Frontier experiment, but they form  a more
diverse set than those of the Energy Frontier. Our  survey found that
there is significant overlap in different experiments' needs. Sharing of
resources across experiments, as in the Open Science Grid, is a first step
in addressing peak computing resource needs.  Continued coordination of
software development between these experiments will allow for efficiently
developed coding infrastructure.  Leveraging the data handling experience and
expertise of the Energy Frontier experiments for the diverse Intensity
Frontier experiments would significantly improve their ability to reconstruct
and analyze data.

{\bf Cosmic Frontier} experiments will greatly expand their storage needs
with the start of new surveys and the development of new instruments.
Current data sets are about 1 PB, and the total data set is expected to be
about 50 PB in ten years. Beyond that, in 10--20 years data will be
collected at the rate of 400 PB/yr. On the astrophysics and cosmology
theory side, some of the most challenging simulations are being run on
supercomputers. 
Current allocations for this effort are approximately 200M core-hours annually.
Very large simulations will require increasing computing
power. Comparing simulations with observations will play a crucial role in
interpretating experiments, and simulations are also needed to help design
new instruments. There are very significant challenges in dealing with new
computer architectures and very large data sets, as described above.
Growing archival storage, visualizing simulations, and allowing public
access to data are also issues that need attention.

{\bf Accelerator science} is called on to simulate new accelerator designs
and to provide near-real-time simulation feedback for accelerator
operation. 
Research into new algorithms and designs has the potential to bring new ideas and
capabilities to the field.
It will be necessary to include additional physics in codes and
to improve algorithms to achieve these goals. Production runs can use from
10K to 100K cores. Considerable effort is being expended to port to new
architectures, especially to address the real-time requirements.

{\bf Lattice gauge theory} calculations rely on national supercomputer
centers and hardware purchased for the USQCD Computing Project. Allocations
at supercomputer centers have exceeded 500 M core-hrs this year, and
resource requests will go up by a factor of 50 by the end of this decade.
This program provides essential input for interpretation of a number of
experiments, and increased precision will be required in the future. For
example, the $b$ quark mass and the strong coupling $\alpha_s$ will need to
be known at the 0.25\% level, a factor of 2 better than now, to compare
upcoming ILC Higgs observations with SM predictions. Advances
in the calculation of hadronic contributions to the muon's anomalous
magnetic moment will be needed
for interpretation of the planned experimental measurement at Fermilab.

{\bf Perturbative QCD} is essential for theoretical understanding of
collider physics rates.  Experts in perturbative QCD computation 
ported codes to the HPC centers at NERSC and
OLCF, and to the Open Science Grid. They have also been
benchmarking GPU codes and finding impressive speed-up over a single core.
A repository of codes has been established at NERSC.  A long-term goal is
to make it easy for experimentalists to use these codes to compute Standard
Model event  rates for the processes they need.

The {\bf Distributed computing and facilities infrastructures} subgroup
looked at the growth trends in distributed resources as provided by the
Open Science Grid, and the national high performance computing 
centers. Most of the computing by experiments is of the HTC type, but HPC
centers could be used for specific work flows. Using existing computing
centers could save smaller experiments from large investments in hardware
and personnel. Distributed HTC has become important in a number of science
areas outside particle physics, but particle physics 
 is still the biggest user and must continue to
drive the future computing development. HPC computing needs for theoretical
physics will require an order of magnitude increase in capacity and
capability at the HPC centers in the next five years, and two orders of
magnitude in the next ten years.

The {\bf Networking} subgroup considered the implications of distributed
computing on network needs, required R\&D and engagement with the National
Research and Education Networks (which carries most of our
traffic). The group formulated a 
number of research questions that need to be answered
before 2020. Expectations of network performance should be raised so that
planning for network needs is on par with that for computing and storage.
The gap between peak bandwidth and delivered bandwidth should be narrowed.
Wide-area network performance should  not be an
insurmountable bottleneck in the next five to ten years as long as
investments in higher performance links continue. However, there is
uncertainty as to whether network costs will drop at the same rate as they
have  in the past.

The {\bf Software development,  personnel and training} subgroup
proposed a
number of recommendations to implement three main goals. The first goal is
to use software development strategies and staffing models that result in
software that is more widely useful to the HEP community. The second goal is to
develop and support software that will run with optimal efficiency on
future computer architectures. The third goal is to ensure that developers
and users have the training necessary to deal with the increasingly complex
software environments and computing systems that will be used in the future.

The {\bf Storage and data management} subgroup found that storage continues
to be a cost driver for many experiments. It is necessary to manage the
cost to optimize the science output from the experiment. Tape storage
continues to be relatively inexpensive and should be utilized more within
the storage hierarchy. 
Disk storage is likely to increase relatively slowly  in capacity per
unit cost,  due
to a shrinking consumer market and technology barriers.
Operating distributed data management systems can be costly for
experiments, and continued R\&D in this area would benefit a number of 
experiments.

To summarize, the challenging resource needs for the planned and proposed
physics programs require efficient and flexible use of all resources. HEP
needs both distributed HTC and HPC. Emerging experimental programs might
consider a mix to fulfill demands. Programs to fund these resources need to
continue. Sharing and opportunistic use help address resource needs, from
all tiers of computing, eventually including commercial providers. There is
increasing need for data-intensive computing in traditionally
computation-intensive fields, including at HPC centers, for data analytics,
combining simulations and observational data, etc.

In order to satisfy our increasing computational demands, the field needs
to make better use of advanced computing architectures. With the need for
more parallelization, the complexity of software and systems continues to
increase, impacting architectures for application frameworks, workload
management systems, and also the physics code. We must develop and maintain
expertise in all areas of the field, and re-engineer frameworks, libraries, and
physics codes. Unless corrective action is taken to enable us to take full
advantage of the new hardware architectures, we could be frozen out of 
cost-effective computing solutions within 10 years. There is a large
code base that needs to be re-engineered, and we currently do not have
enough people trained to do it.

The continuing huge growth in observational and simulation data drives the
need for continued R\&D investment in data management, data access methods,
and networking. Continued evolution of data management and storage
systems will be needed in order to take advantage of new network
capabilities, ensure efficiency and robustness of the global data
federations, and to contain the level of effort needed for operations.
Significant challenges with data management and access remain, and research
into these areas could continue to bring benefit across the frontiers.  We
expect solutions that will be based on content delivery approaches, dynamic
data placement, and remote data access.

Network reliability is essential for data-intensive distributed computing.
Emerging network capabilities and data access technologies improve our
ability to use resources independent of location. This will enable use of
diverse compute resources.  These include dedicated facilities, university computing
centers, and opportunistic use of shared resources between PIs.
They will expand to commercial clouds and  eventually also make use of
 leadership-class HPC
centers relevant for data-intensive computing. The computing models should
treat networks as a resource that needs to be managed and planned for.

Computing will be essential for progress in theory and experiment over the
next two decades. The field continues to learn how to do more science with
constrained resources, requiring us to be more flexible and perhaps tolerate
higher levels of risk. The advances in computer hardware that we have seen
in the past may not continue at the same rate in the future. The issues
identified in this report require continuing attention. Addressing them
will increase efficiency, reduce costs, and enable us to meet the
experimental and theoretical goals identified through the Snowmass process.

\section{Communication, Education, and Outreach}
\label{sec:ceo}

Broad societal support for particle physics research will be required to achieve the 
many scientific and technological goals identified by the U.S. particle physics 
community through the Snowmass process. Building and sustaining this support
 will require the particle physics community to unite behind a common plan
that emerges from the Snowmass/P5 process and to communicate
enthusiasm for the future of the field and its societal impacts to a wider
audience of policy makers, opinion leaders, scientists in other fields,
educators, and students. 

Federally supported research in particle physics and related fields has led to
an impressive list of Nobel Prize-winning discoveries:  the first detailed
study of the cosmic microwave background, the discovery of neutrino masses and
mixing (and earlier work on solar neutrinos many years earlier), the discovery
of the accelerating expansion of the Universe, the understanding of  the strong force,
and the discovery that the CKM matrix explains CP violation. The most 
recent Nobel Prize was awarded for the Higgs boson, whose discovery in 2012
was made possible by scientific talent, technology and
leadership from the United States. 

The American public is fascinated by these discoveries, and by the full breadth of current and future particle physics projects. The saga of the Large Hadron Collider and the Higgs boson discovery reached audience levels unprecedented for a particle physics event. Public lectures and other events on particle physics topics draw crowds. Milestones, discoveries and even proposals for projects in particle physics routinely make headlines. 

Translating this public excitement into greater support for the field requires existing communication, education and outreach (CE\&O) activities to be augmented and enhanced.  We need national coordination and training, additional resources,  and the commitment by the particle physics community to convey consistent, 
coherent and compelling messages
 about the importance of particle physics research and its value to society.

Many individuals, groups, and institutions in the U.S. particle physics community already 
reach out to members of the public, decision makers, teachers, and students through a 
wide variety of effective activities. However, there is room for improvement
 in the nationwide coordination of these activities, in the mobilization of the 
entire U.S. community to take part in the activities, and in efforts to use varied
 activities to convey consistent and compelling messages to stakeholders.

The following is a survey of existing CE\&O activities targeted at four audiences: policy makers and opinion leaders, scientists in other fields, the general public, and educators and students.

{\bf For policy makers and opinion leaders:}
The U.S. particle physics community engages in a number of efforts to build support for
 research among policy makers and opinion leaders. User groups make annual visits to 
Washington, D.C. 
and, with scientific societies, conduct email and letter-writing campaigns at key points
 in the budget cycle. 
Scientists participate in Washington, D.C., events organized by the American Association 
for the Advancement of Science (AAAS), the American Physical Society (APS), the National 
User Facilities Organization (NUFO), and other organizations. Scientific and industrial societies, 
national laboratories, and individual scientists engage in direct advocacy with legislators. 
Scientists and media relations professionals at universities and labs work 
 to place particle physics stories and physicists in influential media outlets.

{\bf For the scientific community:}
Past and current outreach activities of the particle physics community targeted at colleagues in the broader science community include colloquia and seminars at university departments and at national laboratories, and plenary sessions at APS and AAAS meetings. Particle physicists publish their results and pedagogical review articles in journals such as {\it Science} and {\it Nature}. They write articles for  {\it Scientific American}, {\it Popular Science} and similar magazines and for  online forums and science blogs, publish popular science books, and write reports commissioned by labs and agencies.  Examples of the last category include
 {\it Quantum Universe} and {\it Discovering the Quantum Universe}, prepared by HEPAP for the DOE and NSF. 

{\bf For the general public:}
Existing activities that reach the general public are extremely broad but have 
 varying levels of support. The single most common activity is public talks. Scientists frequently participate in open houses and related events such as science festivals, lab and department tours, physics shows, alumni weekends, and workshops for the public. They contribute to external publications and shows by writing magazine articles and op-ed pieces in newspapers, participating as consultants to radio and television programs and movies, and working with the news media. They produce outreach materials such as books, brochures, posters, web-based materials, and multimedia products.  They engage in social media activities such as blogs, Facebook pages, Twitter feeds, and the creation of YouTube videos.

{\bf For educators and students:}
The particle physics community in the United States and abroad has succeeded in increasing student interest and achievement in STEM fields, including particle physics, through a variety of efforts. The 2013 APS Excellence in Physics Education Award was presented to leaders of University of Illinois undergraduate physics education research. The QuarkNet long-term teacher development program has changed how many teachers view science and education by putting cosmic-ray detectors, online analysis tools, and LHC data into their hands. Netzwerk Teilchenwelt adapted the QuarkNet model for students, teachers, and physicists in Germany. The International Particle Physics Outreach Group (IPPOG) sponsored 161 master classes in 37 countries in 2013, including 29 masterclasses in 9 countries in the Fermilab-based portion of the program. National laboratories run successful long-term education programs, many with particle physics content and partnerships with particle physics groups.  The Contemporary Physics Education Project brings together physicists and educators to develop wall charts, posters, websites and activities.

As a result of the Snowmass process, the community has recognized that more physicists must engage in CE\&O activities.  The quality and coordination of such activities must be improved 
in order to  increase public support for the field, develop the next generation of physicists, and ensure scientifically literate and engaged citizens.

A survey of 641 members of the particle physics community conducted in the spring of 
2013 indicated that while about 60\% of physicists engage in outreach to the general 
public and 50\% reach K-12 teachers or students, only 30-35\% engage in activities
 targeted to scientists in other fields or policy makers.
The survey also identified the greatest barriers to participation in CE\&O
activities, including lack of time, little reward in career advancement, and a lack of resources to communicate the broader societal impacts of particle physics research. 

At the Snowmass meeting, a number of prominent voices called for renewed commitment to CE\&O:
\begin{itemize}
   \item ``We must educate our representatives in Congress, our fellow citizens, the business community and the scientific agencies.''  ---  D. Gross
   \item ``You are underselling yourselves\textellipsis you are technology incubators for other fields of science.'' --- R. Roser 
   \item ``The media missed the substantial impact of the U.S. on the Higgs discovery.''  ---  J. Incandela
   \item ``You need to appeal to varied stakeholders to convince them that you do valuable science with a sensible plan. Illustrate the benefits of particle physics to society.''  ---  G. Blazey
\end{itemize}

The CE\&O group developed the  following goals, strategies, and recommendations with input from particle physicists and education and outreach professionals. The recommendations support a proactive, coordinated CE\&O effort from the entire U.S. particle physics community. 

As overarching goals for U.S. particle physics communication, education and outreach, we recommend:
\begin{enumerate}
\item Ensuring that the U.S. particle physics community has the resources necessary to conduct research and maintain a world leadership role.
\item  Ensuring that the U.S. public appreciates the value and excitement of particle physics.
\item  Ensuring that a talented and diverse group of students enters particle physics and other STEM careers, including science teaching.
\end{enumerate}

We  recommend five-year CE\&O implementation recommendations that cut across all  audiences:
\begin{enumerate}
\item Augment existing efforts with additional personnel and resources dedicated to nationwide coordination, training and support.
\item Develop a comprehensive central communication, education and outreach resource for physicists, with initial content available before the end of the 2013/2014 P5 process.
\item  Provide communication training to the U.S. particle physics community.
\item  Work with DPF and HEPAP to develop a sustainable process for collecting statistics on workforce development and technology transfer and with APS to investigate a U.S. economic impact study for physics research that includes particle physics.
\end{enumerate}

We further recommend strategies for specific audiences:

{\bf For policy makers and opinion leaders:}
(1)  Empower and enable members of the particle physics community to communicate and advocate coherently, consistently, and effectively on behalf of their science.
(2) Develop an enduring process to track, update, and disseminate statistics on the impact of particle physics on society.  (3) Put informed third-party advocates to work raising the profile of and informing key stakeholders about the importance of particle physics, physics, and basic science to the United States.

{\bf For the scientific community:}
(1)  Foster more dialog and understanding between subfields of science.
(2)  Identify areas of common cause and unite in support of them.
(3)  Develop consensus in our field that we need to prioritize, buy into the mechanism of prioritization, and then support the resulting plan.

{\bf For the general public:}
(1)  Engage the public in a wide range of outreach activities. 
(2)  Make the public aware of direct and indirect applications of research, both historical and potential. 
(3) Communicate the role and stories of U.S. physicists in particle physics, particularly in major discoveries and in the context of our international collaborations.

{\bf For educators and students:}
(1)  Directly engage with students and educators.  Invite educators and students into our unique community. 
 (2)  Offer long-term professional development and training opportunities for educators (including pre-service educators), aligned with current and appropriate standards and enabling educators to explore best-practice teaching methods. Make an effort to collaborate with local schools of education whenever possible.
   (3)  Create learning opportunities for students of all ages, including classroom, out-of-school and online activities that allow students to explore particle physics to construct their own understanding and develop the skills and habits of mind necessary to perform research.

{\bf Resources:}
An overarching recommendation that supports all goals and strategies is the augmentation of existing efforts with additional personnel and resources dedicated to nationwide coordination, training, and support for particle physics education, outreach, and communication activities. Such a team would enhance existing efforts and spearhead new initiatives, such as the development of a comprehensive central communication, education, and outreach resource for physicists, the development of sustainable methods to collect statistics on workforce development and technology transfer, materials designed to inform the public about direct and indirect applications of particle physics, the creation of professional development opportunities for educators, and new learning opportunities for students of all ages. 





\section{Conclusion}

Here we recapitulate the Conclusion given previously in our Executive Summary.
With the completion of the Standard Model, particle physicists now turn their
attention to still deeper questions about the nature of matter and the
constituents of the universe.
This report proposes an ambitious array of new experiments.  We
consider it realistic to carry out these experiments through a long-term plan
and through global partnerships.  Particle physicists have been the pioneers of
large-scale scientific projects.  We have constructed facilities of
unprecedented scale, including the Tevatron and the Large Hadron Collider,
through decades-long programs requiring world-wide collaboration.  These led
to discoveries that are the foundation of our current success.

Several strategic goals have emerged from the Snowmass study.

\begin{itemize}

\item Probe the highest possible energies and distance scales with the
existing and upgraded Large Hadron Collider and reach for even higher
precision with a lepton collider; study the properties of the Higgs boson
in full detail.

\item Develop technologies for the long-term future to build multi-TeV
lepton colliders and 100 TeV hadron colliders.

\item Execute a program with the U.S. as host that provides precision tests
of the neutrino sector with an underground detector; search for new physics
in quark and lepton decays in conjunction with precision measurements of
electric dipole and anomalous magnetic moments.

\item Identify the particles that make up dark matter through complementary
experiments deep underground, on the Earth's surface, and in space, and
determine the properties of the dark sector.

\item Map the evolution of the universe to reveal the origin of cosmic
inflation, unravel the mystery of dark energy, and determine the ultimate
fate of the cosmos.

\item Invest in the development of new, enabling instrumentation and
accelerator technology.

\item Invest in advanced computing technology and programming expertise
essential to both experiment and theory.

\item Carry on theoretical work in support of experimental projects and to
explore new unifying frameworks.

\item Invest in the training of physicists to develop the most creative
minds to generate new ideas in theory and experiment that advance science
and benefit the broader society.

\item Establish a nationally coordinated communication, education and outreach
  effort, supported by a dedicated team, to convey the excitement and
  value of our field to others.

\end{itemize}

In pursuit of these projects, we have developed a community that links
together scientists from all regions of the world pursuing common goals.  Our
community is ready and eager to carry out the next steps in humankind's
quest to understand the basic workings of the universe.

\end{document}